\documentclass[
  aps,
  prl,
  reprint,
  nofootinbib,
  longbibliography
]{revtex4-2}

\usepackage[T1]{fontenc}
\usepackage{graphicx}
\usepackage{amsmath,amssymb,bm,mathtools}
\usepackage{microtype}
\usepackage{float}

\newcommand{\Tr}{\operatorname{Tr}}

\begin{document}

\title{Quantum Interference Enhanced Rectification in Mesoscopic Nanowires}

\author{T. J. Osborne}
\author{B. Braunecker}
\author{H. F. Legg}
\affiliation{School of Physics and Astronomy, University of St Andrews, North Haugh, St Andrews KY16 9SS, United Kingdom}

\date{\today}

\begin{abstract}
Nanoscale nonreciprocal elements are needed to control signal flow in scalable quantum technologies, but existing nonreciprocal components are often bulky. Here, we show that a strongly nonreciprocal current--voltage relation can arise entirely from coherent elastic scattering in a mesoscopic Rashba nanowire with gate-defined scatterers. We first identify the symmetry constraints that enforce reciprocal transmission in a phase coherent conductor even when inversion and time-reversal symmetry are broken. We then show how gates that are capacitively coupled to the leads, together with a magnetic scatterer or transverse subband mixing, can remove these constraints. In the weak-scattering regime, nonreciprocal conductance originates from interference between distinct backscattering paths. In the strong-scattering regime, nonreciprocity can be resonantly enhanced by the hybridization of the transmitting channel with bound states that arise from a higher transverse subband or spin-split band. In this regime, we find that a bias-induced change in the gate potentials of about one percent can produce an order-unity diode response, i.e., electrons are transmitted almost as in a clean wire for one bias polarity, whereas for the other polarity the current is strongly suppressed.
\end{abstract}

\maketitle

As qubit platforms scale to larger processors, elements with nonreciprocal transmission, such as isolators and circulators, are needed to protect quantum devices from noise and reflected signals~\cite{abdo2019active,Chapman2017,Barzanjeh2017,Bernier2017,Abdo2021,Rosenthal2021}. Existing implementations are often bulky or rely on temporal modulation or Josephson junctions~\cite{nagulu2020non}. Such ingredients can introduce dissipation and additional noise, complicating their integration near fragile quantum elements. Understanding how to realize nonreciprocal effects in low-temperature nanoscale devices is therefore a key requirement for scalable quantum hardware.

\begin{figure}[!t]
    \centering
    \includegraphics{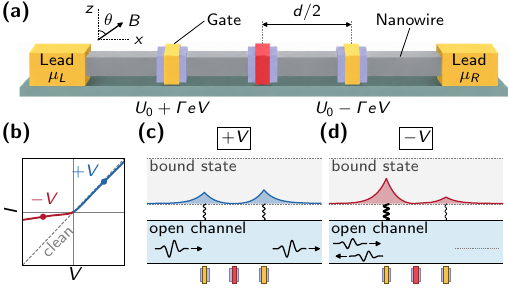}
    \caption{\textbf{Minimal device for interference-enhanced rectification in mesoscopic nanowires.} (a) Rashba nanowire along $x$ in a magnetic field $B$ in the $xz$ plane, at an angle $\theta$ to the $z$ axis. Two outer gates separated by $d$ acquire bias-dependent strengths $U_0\pm\Gamma eV$. The central red element is a magnetic scatterer in the 1D model or a bias-independent scalar gate in the quasi-1D model with two transverse subbands. (b) Schematic current--voltage characteristic of the device compared with that of the clean wire (gray), with dots at the operating bias $\pm V$. (c,d) Mechanism at $\pm V$. The gates bind a state in a closed channel (bound state), namely the second transverse subband (quasi-1D) or the upper spin-split band (1D), which couples to the open channel.}
    \vspace{-10pt}
    \label{fig:devices}
\end{figure}

One route to realizing such devices is to exploit nonlinear transport effects~\cite{Tokura2018,NagaosaYanase2024}, in which the current $I$ through a conductor responds asymmetrically to the applied bias voltage $V$, i.e., $I(V)\neq-I(-V)$, a phenomenon known as rectification or the diode effect [see Fig.~\ref{fig:devices}(b)]. Intrinsic nonlinear responses such as electrical magnetochiral anisotropy have been observed in noncentrosymmetric conductors~\cite{Rikken2001,Yokouchi2017}, polar Rashba semiconductors~\cite{Ideue2017,Li2021}, oxide interfaces~\cite{Yoo2019}, topological semimetals~\cite{Wang2022}, and topological insulators~\cite{Legg2022,He2018}. However, these responses are commonly described within semiclassical diffusive transport~\cite{RikkenWyder2005,Morimoto2016,Dantas2023}. In micro- and nanoscale devices comparable in size to the phase-coherence length, elastic scattering and quantum interference instead replace dissipative momentum relaxation as the central ingredients of rectification~\cite{zou2024nonreciprocal,Leturcq2006,Angers2007,HernandezLewenkopf2009}.

Semiconductor and topological insulator nanowires~\cite{vanWeperen2013,Heedt2016,Kammhuber2016,Li2016,Kammhuber2017,Peng2010,Dufouleur2013,Cho2015,Rossler2023,Legg2022} provide platforms for phase-coherent, few-mode transport. In semiconductor nanowires, for instance InAs or InSb, broken inversion symmetry results in a strong Rashba spin--orbit coupling~\cite{Manchon2015,vanWeperen2015}. In topological insulator nanowires, inversion symmetry can instead be broken by a gate~\cite{Legg2022}. In both cases, an externally applied magnetic field additionally breaks time-reversal symmetry. The same combination of broken symmetries makes such nanowires not only promising for nonreciprocal transport~\cite{angehrn2024relations}, but also important for spintronics~\cite{DattaDas1990,NadjPerge2010} and hybrid superconducting devices~\cite{Oreg2010,Lutchyn2010,Mourik2012,Krogstrup2015,Albrecht2016,Gul2018,Alicea2012,Lutchyn2018,Prada2020,legg2022superconducting}. 

In a coherent two-terminal conductor, however, breaking inversion and time-reversal symmetry is not sufficient for nonreciprocal transport, since the unitarity of the scattering matrix and residual unitary or antiunitary symmetries impose additional constraints on rectification~\cite{Buttiker1988,Buttiker1993,ChristenButtiker1996,SanchezButtiker2004,SpivakZyuzin2004}. Rectification in mesoscopic conductors has been demonstrated in devices with an asymmetric geometry, for instance ballistic rectifiers~\cite{Song1998,Fleischmann2002} and electron ratchets~\cite{Linke1999}. Additionally, electron--electron interactions can generate rectification from asymmetric scatterers~\cite{ChristenButtiker1996,braunecker2005rectification,FeldmanScheidlVinokur2005}. In each case, however, the asymmetry that distinguishes the two bias directions is built into the device geometry or supplied by interactions. 

In this paper, we show that phase-coherent interference between bias-responsive scatterers provides a route to strong rectification in quasi-ballistic Rashba nanowires. We identify the symmetries that enforce reciprocal transmission in coherent two-terminal transport and show that capacitive coupling between the gates and the leads can supply the bias dependence required to cause nonreciprocal effects. For minimal devices that evade these constraints (see Fig.~\ref{fig:devices}), weak scatterers rectify through interference between distinct backscattering paths, as we show analytically. Beyond this regime, we find numerically that states from a higher transverse subband or spin-split band that are bound by the gates can resonantly amplify a bias-induced change of about $1\%$ in the gate potentials into a difference of about a factor of ten between the currents for the two bias directions [see Fig.~\ref{fig:devices}(b--d)].

\emph{Symmetry constraints}---We apply a symmetric bias $V$, i.e., the electrochemical potentials of the left and right leads are $\mu_L=eV/2$ and $\mu_R=-eV/2$ (see Fig.~\ref{fig:devices}). We choose a symmetric bias so that, in a mirror-symmetric device, the bias shifts the two gate potentials only in opposite directions [Eq.~\eqref{eq:bias-dependent-potentials}], whereas grounding one lead would add a common shift of both gates proportional to $V$, a separate source of bias dependence that we do not consider. In a mirror-symmetric device, the potential at the center then remains at its equilibrium value, as we show in the Supplemental Material (SM)~\cite{SM}. As such, rectification cannot arise merely from the choice of grounded lead~\cite{ChristenButtiker1996}. Coherent two-terminal transport is described by the Landauer formula~\cite{Landauer1957,Landauer1970,Buttiker1986}, which for the electron current reads
\begin{equation}
 I(V)=\frac{e}{2\pi \hbar}\int dE\,\mathcal T(E,V)\,[f_L(E,V)-f_R(E,V)],
 \label{eq:landauer-main}
\end{equation}
where $e>0$ is the elementary charge, $\mathcal T(E,V)$ is the transmission probability at energy $E$, and $f_\alpha(E,V)=\{\exp[(E-\mu_\alpha)/k_{\mathrm B}T]+1\}^{-1}$, with $\alpha\in\{L,R\}$, are the Fermi--Dirac distributions of the reservoirs. The integral runs over all energies, measured from the equilibrium chemical potential, but for $k_{\mathrm B}T\ll e|V|$ only the bias window $|E|\lesssim e|V|/2$ contributes. Since the occupation difference $\Delta f(E,V)=f_L(E,V)-f_R(E,V)$ is odd under bias reversal, $V\to-V$, the rectified current is controlled by the odd-in-bias part of the transmission, $\mathcal T_{\rm odd}(E,V)=[\mathcal T(E,V)-\mathcal T(E,-V)]/2$. In particular,
\begin{equation}
 I(V)+I(-V)=\frac{e}{\pi\hbar}\int dE\,\mathcal T_{\rm odd}(E,V)\,\Delta f(E,V).
 \label{eq:ieven-main}
\end{equation}
To determine when $\mathcal T_{\rm odd}$ must vanish, we express the transmission through the scattering matrix of the propagating lead modes~\cite{Buttiker1988,beenakker1997random},
\begin{equation}
 S(E,V)=\begin{pmatrix} r & t' \\ t & r' \end{pmatrix},
 \label{eq:S-matrix}
\end{equation}
where $r$ and $t$ ($r'$ and $t'$) are the reflection and transmission matrices for incidence from the left (right) lead. The transmission is $\mathcal T=\Tr(t^\dagger t)$ for incidence from the left and $\mathcal T'=\Tr(t'^\dagger t')$ for incidence from the right. For flux-normalized modes, $S$ is unitary, which encodes current conservation~\cite{Buttiker1988} and enforces that the transmission is the same for incidence from either lead at a fixed bias, $\mathcal T'(E,V)=\mathcal T(E,V)$. The transmission is also reciprocal,
\begin{equation}
 \mathcal T(E,-V)=\mathcal T(E,V),
 \label{eq:no-go-main}
\end{equation}
if a unitary or antiunitary transformation maps the complete forward-bias problem, including the leads, onto the reverse-bias problem. In other words, the transmission is reciprocal if an electron cannot distinguish the two bias directions~\cite{Buttiker1988,ChristenButtiker1996,SanchezButtiker2004,SpivakZyuzin2004} (see SM for details~\cite{SM}). In particular, if the reverse-biased device is the mirror image or time-reversed copy of the forward-biased one, an electron incident from the left at $-V$ is transmitted with the same probability as one incident from the right at $+V$, hence $\mathcal T(E,-V)=\mathcal T'(E,V)=\mathcal T(E,V)$. We emphasise that transformations that enforce reciprocity can exist even when both time-reversal and inversion symmetry are broken.

\begin{figure}[t]
 \includegraphics{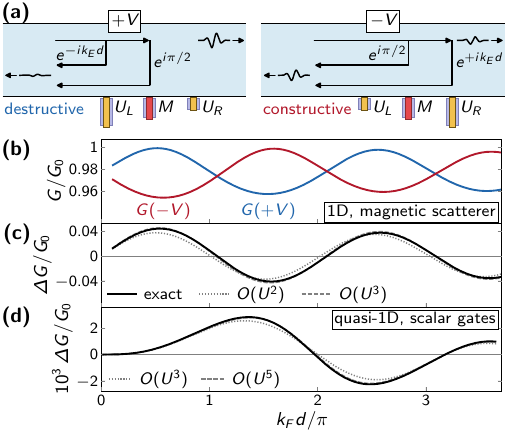}
 \caption{\textbf{Weak-scattering rectification and its Born-interference origin.}
(a) First-order scalar and magnetic backscattering paths in the 1D model for an electron incident from the left, which interfere destructively at $+V$ and constructively at $-V$, drawn for $k_Ed\approx\pi/2$ and $M<0<U_0$; the taller outer gate is the stronger one.
(b,c) Directional conductances and diode effectiveness $\Delta G/G_0$ [Eq.~\eqref{eq:eta}] for the 1D model and (d) $\Delta G/G_0$ for the quasi-1D model with three weak scalar gates (see SM for details~\cite{SM}), numerically exact (solid) and in the Born expansion to the indicated order (dotted, dashed). Both calculations use $m^*=0.014m_e$, $\alpha_R=0.5\,\mathrm{eV\,\mathring{A}}$, $T=25\,\mathrm{mK}$, and bias-dependent gates, $\Gamma eV/U_0\approx1$.}
\label{fig:weak-born-mechanisms}
\end{figure}

\emph{Bias-dependent scatterers}---A bias-odd transmission, which Eq.~\eqref{eq:ieven-main} requires for rectification, can arise when the gates that define the scattering potentials are capacitively coupled to the source and drain reservoirs~\cite{Buttiker1993,ChristenButtiker1996}. A single off-center gate then already rectifies, although for realistic couplings less strongly than the resonant quasi-1D device below and only in a narrow range of gate strengths (see Fig.~S2 of the SM~\cite{SM}). To exploit interference, we instead place two gates symmetrically about the center of the device, such that they exchange their capacitances to the two leads under inversion. If the gates are electrically floating, their charges remain fixed and the bias shifts their potentials by equal and opposite amounts (see SM~\cite{SM}). In the linear electrostatic regime, this gives
\begin{equation}
    U_{L/R}(V)=U_0\pm\Gamma eV,\qquad U(x,-V)=U(-x,V),
    \label{eq:bias-dependent-potentials}
\end{equation}
where $U(x,V)=U_L(V)\delta(x+d/2)+U_R(V)\delta(x-d/2)$ describes the gates as point-like potentials, $U_0$ is their zero-bias strength, and $\Gamma$, which has units of length, parametrizes the electrostatic coupling. This description requires gates narrower than both the Fermi wavelength and the decay length of the bound states that produce the resonances discussed below. Under bias reversal, the two gate potentials are interchanged, which alone cannot produce rectification. If spatial inversion $\mathcal P$, which takes $x\to-x$, combined with a spin operation or complex conjugation, leaves the clean nanowire invariant, this bias-independent symmetry maps the scattering problem at forward bias onto that at reverse bias, such that Eq.~\eqref{eq:no-go-main} enforces reciprocal transmission. Rectification therefore requires an additional ingredient that removes this map, e.g., a magnetic scatterer in a single-subband (1D) model, or the coupling between transverse subbands in a two-subband (quasi-1D) model.

\emph{1D quantum interference}---We first retain a single transverse subband of the standard Rashba nanowire model~\cite{Oreg2010,Lutchyn2010}, described by the clean-wire Hamiltonian
\begin{equation}
H_1(k)
=
\left(
\frac{\hbar^2 k^2}{2m^\ast}-\mu
\right)\sigma_0
+\alpha_R k\,\sigma_y
+\Delta_x\sigma_x,
\label{eq:single-band-hamiltonian}
\end{equation}
where $k$ is the wavevector along the nanowire, $m^*$ the effective mass, $\mu$ the chemical potential, $\alpha_R$ the Rashba spin--orbit coefficient, and $\Delta_x$ the Zeeman energy due to a magnetic field along the nanowire [$\theta=\pi/2$ in Fig.~\ref{fig:devices}(a)]. The identity $\sigma_0$ and the Pauli matrices $\sigma_{x,y,z}$ act in spin space.

The simplest geometry that could resonantly amplify the response consists of two scalar gates at $x=\pm d/2$. However, since Eq.~\eqref{eq:bias-dependent-potentials} maps this gate profile onto itself under inversion and bias reversal, the device possesses the lead-exchanging antiunitary symmetry $\mathcal P\sigma_x\mathcal K$ (see SM~\cite{SM}), where $\sigma_x$ is the spin component along the field and $\mathcal K$ is complex conjugation. We can break this symmetry by placing a local magnetic scatterer $M\sigma_z$ at the origin [see Fig.~\ref{fig:devices}(a)].

The two bias polarities then differ because the reflection from the magnetic scatterer is shifted by $\pi/2$ relative to that from a gate, since the matrix element of $\sigma_z$ between the counterpropagating lower-band spinors is $-i\alpha_Rk_E/\Delta_x$ times their overlap (see SM~\cite{SM}). Here, $k_E>0$ is the wavevector of the right-moving lower-band state at energy $E$. Bias reversal swaps which outer gate is stronger and thus the sign of the dominant scalar propagation phase $e^{\mp ik_Ed}$. For $k_Ed\approx\pi/2$, the dominant paths therefore interfere destructively at $+V$ and constructively at $-V$, leading to rectification [see Fig.~\ref{fig:weak-born-mechanisms}(a)].

In the single-open-channel regime, in which only the lower band $E_-(k)$ of Eq.~\eqref{eq:single-band-hamiltonian} has states at energy $E$, at its two Fermi points $\pm k_E$, the leading bias-odd transmission arises from interference between the first-order scalar and magnetic backscattering amplitudes and is given by
\begin{equation}
\mathcal T_{\rm odd}^{(2)}(E,V)
=
-\frac{4M\Gamma eV}
{\hbar^{2}\lvert v(k_E)\rvert^{2}}
\frac{\alpha_R k_E \Delta_x}
{(\alpha_R k_E)^{2}+\Delta_x^{2}}
\sin(k_E d),
\label{eq:continuum-bs-odd}
\end{equation}
where the superscript denotes the order in the scatterer strengths and $v(k_E)=\hbar^{-1}\partial_kE_-(k)|_{k=k_E}$ is the group velocity (see SM~\cite{SM}). This contribution therefore requires spin--orbit coupling, a finite Zeeman energy $\Delta_x$, both scalar and magnetic scattering, and a bias-responsive potential. Furthermore, the sinusoidal dependence on $k_Ed$ is a direct signature of coherent interference between the spatially separated scatterers [see Fig.~\ref{fig:weak-born-mechanisms}(b,c)].

We will quantify the diode response by the effectiveness
\begin{equation}
 \frac{\Delta G}{G_0}=\frac{I(+V)+I(-V)}{G_0|V|}=\frac{G(+V)-G(-V)}{G_0},
 \label{eq:eta}
\end{equation}
where $G(\pm V)=\pm I(\pm V)/V$, for $V>0$, are the directional conductances and $G_0=e^2/2\pi\hbar$. A value $|\Delta G/G_0|\sim1$ indicates that the even-in-bias current is comparable to the ballistic current of a single spin-resolved channel. We note that the effectiveness is a stricter figure of merit than the diode efficiency $\eta=[G(+V)-G(-V)]/[G(+V)+G(-V)]$, which can be large even when both currents are small.

In the weak-scattering regime, the directional conductances remain close to the ballistic value. For the parameters of Fig.~\ref{fig:weak-born-mechanisms}, they also oscillate in antiphase as a function of $k_Fd/\pi$, where $k_F=k_{E=0}$, which generates an effectiveness of a few percent, as Eq.~\eqref{eq:continuum-bs-odd} already captures. The third-order Born expansion closely follows the numerically exact result, which includes all orders of scattering and all evanescent modes (see SM~\cite{SM}).
\begin{figure*}[t!]
    \centering
    \includegraphics{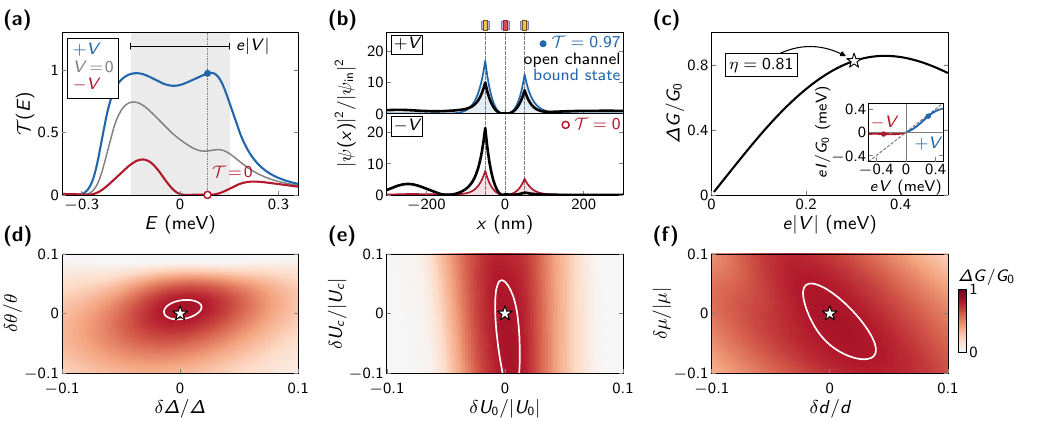}
    \caption{\textbf{Resonantly enhanced quasi-1D rectifier.}
(a) Transmission of the three-gate device at $+V$ (blue), $V=0$ (gray), and $-V$ (red), with the bias window shaded and the exact zero at $E_0$ for $-V$ marked by the open circle. (b) Densities $|\psi(x)|^2$ of the scattering states at $E_0$ for incidence from the left in the lowest (open channel, black) and second subband (bound state, colored), relative to the uniform density $|\psi_{\rm in}|^2$ of the incident open-channel wave; markers as in (a). (c) Diode effectiveness $\Delta G/G_0$ [Eq.~\eqref{eq:eta}] versus bias magnitude, with the diode efficiency $\eta$ at the operating point (star). Inset: current--voltage characteristic of the device, with dots at $\pm V$, and of the clean wire (dashed). (d--f) $\Delta G/G_0$ under simultaneous deviations of the Zeeman energy and field angle, $(\Delta,\theta)$, the outer and central gate strengths, $(U_0,U_c)$, and the gate separation and chemical potential, $(d,\mu)$, from the operating point (stars). White contours indicate $\Delta G/G_0=0.8$. We use $m^\ast=0.014m_e$, $\alpha_R=0.5\,\mathrm{eV\,\mathring{A}}$, $W_y=200\,\mathrm{nm}$, $T=25\,\mathrm{mK}$, and $|\Gamma eV/U_0|\simeq1\%$, with the other parameters listed in the SM~\cite{SM}.}
    \label{fig:two-band-continuum-strong}
\end{figure*}

\emph{Quasi-1D quantum interference}---Purely scalar gates with no magnetic scatterer, desirable for nanoscale devices, become sufficient for rectification when the coupling between the transverse subbands of a realistic nanowire is included and the field has components both along the nanowire and along $z$ ($0<\theta<\pi/2$), as described by the clean-wire Hamiltonian of the two lowest subbands
\begin{multline}
H_2(k)
=
\left(
\frac{\hbar^2 k^2}{2m^\ast}-\mu
\right)\sigma_0\tau_0
+\alpha_R k\,\sigma_y\tau_0
+\Delta_x\sigma_x\tau_0\\
+\Delta_z\sigma_z\tau_0
+\frac{\Delta_\perp}{2}\sigma_0(\tau_0-\tau_z)
+\lambda_\perp\sigma_x\tau_y,
\label{eq:two-band-continuum-hamiltonian}
\end{multline}
where the identity $\tau_0$ and the Pauli matrices $\tau_{x,y,z}$ act in the space of these subbands, $\Delta_\perp$ is their separation, $\lambda_\perp$ is the transverse Rashba coupling between them, and the Zeeman energy $\Delta$ of a magnetic field $B$ at an angle $\theta$ to the $z$ axis [see Fig.~\ref{fig:devices}(a)] has components $\Delta_x=\Delta\sin\theta$ and $\Delta_z=\Delta\cos\theta$. Both $\Delta_\perp$ and $\lambda_\perp$ are set by the transverse confinement of width $W_y$ (see SM~\cite{SM}). The mixing $\lambda_\perp$, together with a tilted Zeeman field, removes all lead-exchanging forward--reverse maps. For $\Delta_z=0$ ($\Delta_x=0$), however, the map $\mathcal P\tau_z\sigma_x\mathcal K$ ($\mathcal P\sigma_z\mathcal K$) restores reciprocity (see SM~\cite{SM}). Since the scalar gates do not depend on the field, time reversal maps the device at $B$ onto that at $-B$ without exchanging the leads, so $\Delta G$ is even in $B$, unlike magnetochiral anisotropy. Reversing only $\Delta_x$ or only $\Delta_z$, in contrast, changes the sign of $\Delta G$ (see SM~\cite{SM}).

For this model, two bias-dependent scalar gates, i.e., the device of Fig.~\ref{fig:devices}(a) without the central red scatterer, already rectify. However, every scalar gate backscatters with the same spinor overlap between the counterpropagating states of the lowest band, so the transmission remains even in bias at second order in the gate strengths (see SM~\cite{SM}). A central gate of strength $U_c$ at the origin, which is bias independent by symmetry, adds propagation phases associated with both $d$ and $d/2$ [see Fig.~\ref{fig:devices}(a)]. For weak gates, the effectiveness nonetheless remains more than an order of magnitude below that of the 1D model with a magnetic scatterer [see Fig.~\ref{fig:weak-born-mechanisms}(d) and SM~\cite{SM}].

\emph{Resonantly enhanced rectification}---A much larger response emerges beyond the weak-scattering regime, where repeated coherent scattering produces transmission resonances. At a realistic operating point with $e|V|\approx0.3$~meV (see SM~\cite{SM}), one polarity remains close to the clean-wire current, $G(+V)\approx0.9G_0$, whereas the other is suppressed by about a factor of ten [see Fig.~\ref{fig:two-band-continuum-strong}(a) and the inset of Fig.~\ref{fig:two-band-continuum-strong}(c)]. As such, we achieve an effectiveness $\Delta G/G_0=0.82$ and efficiency $\eta=0.81$. The effectiveness stays above $0.8$ for $e|V|$ between about $0.28$ and $0.46$~meV [see Fig.~\ref{fig:two-band-continuum-strong}(c)] and survives percent-level variations of the field, the gate strengths and separation, and the chemical potential [see Fig.~\ref{fig:two-band-continuum-strong}(d--f) and SM~\cite{SM}]. Orbital effects of the field and weak disorder, which we neglect, cannot restore reciprocity but can shift the resonances. With orbital effects included, retuning restores a comparable response (see SM~\cite{SM}). In general, we expect that a device shorter than the elastic mean free path will exhibit the effect~\cite{vanWeperen2013,Kammhuber2016}.

The resonances originate from states bound by the attractive gates, $U_0,U_c<0$, in the second transverse subband. Without transverse coupling, the gates reflect almost every electron in the lowest subband (see SM~\cite{SM}). The coupling $\lambda_\perp$ hybridizes the bound states with the open channel and turns them into broad resonances of Fano type~\cite{Fano1961,Bagwell1990,GurvitzLevinson1993,Miroshnichenko2010} [see Fig.~\ref{fig:two-band-continuum-strong}(a) and SM~\cite{SM}], such that the response requires this coupling and grows approximately as $\lambda_\perp^2$ for weak coupling. The same mechanism in the strong-scattering regime can occur in the 1D model, with a magnetic scatterer, where the upper spin-split band plays the same role as the second transverse subband (see SM~\cite{SM}).

For reverse bias, the transmission vanishes exactly at two energies inside the bias window (see SM~\cite{SM}). The zero at higher energy, $E_0$, is marked by the open circle in Fig.~\ref{fig:two-band-continuum-strong}(a). Since the two zeros lie close together, they pin $\mathcal T(E,-V)$ to small values across the window, such that the integral in Eq.~\eqref{eq:landauer-main} remains small. For forward bias, the transmission instead stays large. These zeros require no fine-tuning, since the antiunitary symmetry $\tau_z\mathcal K$ of the quasi-1D model ensures that transmission zeros lie on the real energy axis or form complex-conjugate pairs~\cite{NockelStone1994,Lee1999}. A small field component along $y$, about $1\%$ of the total, breaks this symmetry but barely changes the response (see SM~\cite{SM}). Unlike the maps behind Eq.~\eqref{eq:no-go-main}, this symmetry does not relate opposite biases, and the zeros enter the window for a sufficiently large bias (see SM~\cite{SM}).

The difference between the two polarities is visible in the scattering state $\psi(x)$ of an electron that arrives from the left at $E_0$ as the open-channel plane wave $\psi_{\rm in}(x)$, which we compute exactly and project onto the two subbands [see Fig.~\ref{fig:two-band-continuum-strong}(b) and SM~\cite{SM}]. Since $|\psi_{\rm in}|^2$ is uniform, we show $|\psi(x)|^2/|\psi_{\rm in}|^2$, which is independent of the normalization of the incident wave and equals one in a clean wire. At $+V$, the second-subband density is spread over both outer gates and the electron is almost fully transmitted. At $-V$, in contrast, the electron is fully reflected at the first gate, where it dwells near the resonance and the open-channel density exceeds $20$ times the incident density. Since $E_0$ is a transmission zero, this density vanishes at the right lead.

\emph{Conclusions}---We have shown that coherent elastic interference can produce a strong, gate-induced diode response in a two-terminal Rashba nanowire, provided that a magnetic scatterer (1D model) or transverse Rashba coupling and a tilted Zeeman field (quasi-1D model) remove the symmetries relating the two bias directions. States bound by the gates in a higher transverse subband or spin-split band can resonantly enhance the nonreciprocity, such that one polarity is transmitted almost ballistically whereas the current for the other is about ten times smaller. For parameters typical of InSb nanowires~\cite{vanWeperen2015}, these results identify few-mode Rashba nanowires as a promising platform for compact, electrically tunable mesoscopic rectifiers.

\begin{acknowledgments}
\textit{Acknowledgements:} T. O. acknowledges the support from the EPSRC under Grant No.~EP/Z535187/1.

\textit{Data availability:} The data that support the findings of this article, together with the code that
generates all figures and numerical results, are openly available on Zenodo~\cite{Zenodo}.

\textit{AI usage:} We used Claude Opus~5.5 and Claude Fable~5.1 for consistency checks, internal review, code improvement, and cross-checks of numerical results.
\end{acknowledgments}

\nocite{Nakatsukasa2018,Nilsson2009}
\bibliography{rectification}

@article{abdo2019active,
  author = {Abdo, B. and Bronn, N. T. and Jinka, O. and Olivadese, S. and C{\'o}rcoles, A. D. and Adiga, V. P. and Brink, M. and Lake, R. E. and Wu, X. and Pappas, D. P. and Chow, J. M.},
  title = {Active protection of a superconducting qubit with an interferometric {Josephson} isolator},
  journal = {Nat. Commun.},
  volume = {10},
  pages = {3154},
  year = {2019},
  doi = {10.1038/s41467-019-11101-3}
}

@article{Chapman2017,
  author = {Chapman, B. J. and Rosenthal, E. I. and Kerckhoff, J. and Moores, B. A. and Vale, L. R. and Mates, J. A. B. and Hilton, G. C. and Lalumi{\`e}re, K. and Blais, A. and Lehnert, K. W.},
  title = {Widely tunable on-chip microwave circulator for superconducting quantum circuits},
  journal = {Phys. Rev. X},
  volume = {7},
  pages = {041043},
  year = {2017},
  doi = {10.1103/PhysRevX.7.041043}
}

@article{Barzanjeh2017,
  author = {Barzanjeh, S. and Wulf, M. and Peruzzo, M. and Kalaee, M. and Dieterle, P. B. and Painter, O. and Fink, J. M.},
  title = {Mechanical on-chip microwave circulator},
  journal = {Nat. Commun.},
  volume = {8},
  pages = {953},
  year = {2017},
  doi = {10.1038/s41467-017-01304-x}
}

@article{Bernier2017,
  author = {Bernier, N. R. and T{\'o}th, L. D. and Koottandavida, A. and Ioannou, M. A. and Malz, D. and Nunnenkamp, A. and Feofanov, A. K. and Kippenberg, T. J.},
  title = {Nonreciprocal reconfigurable microwave optomechanical circuit},
  journal = {Nat. Commun.},
  volume = {8},
  pages = {604},
  year = {2017},
  doi = {10.1038/s41467-017-00447-1}
}

@article{Abdo2021,
  author = {Abdo, B. and Jinka, O. and Bronn, N. T. and Olivadese, S. and Brink, M.},
  title = {High-fidelity qubit readout using interferometric directional {Josephson} devices},
  journal = {PRX Quantum},
  volume = {2},
  pages = {040360},
  year = {2021},
  doi = {10.1103/PRXQuantum.2.040360}
}

@article{Rosenthal2021,
  author = {Rosenthal, E. I. and Schneider, C. M. F. and Malnou, M. and Zhao, Z. and Leditzky, F. and Chapman, B. J. and Wustmann, W. and Ma, X. and Palken, D. A. and Zanner, M. F. and Vale, L. R. and Hilton, G. C. and Gao, J. and Smith, G. and Kirchmair, G. and Lehnert, K. W.},
  title = {Efficient and low-backaction quantum measurement using a chip-scale detector},
  journal = {Phys. Rev. Lett.},
  volume = {126},
  pages = {090503},
  year = {2021},
  doi = {10.1103/PhysRevLett.126.090503}
}

@article{nagulu2020non,
  author = {Nagulu, A. and Reiskarimian, N. and Krishnaswamy, H.},
  title = {Non-reciprocal electronics based on temporal modulation},
  journal = {Nat. Electron.},
  volume = {3},
  pages = {241},
  year = {2020},
  doi = {10.1038/s41928-020-0400-5}
}

@article{Tokura2018,
  author = {Tokura, Y. and Nagaosa, N.},
  title = {Nonreciprocal responses from non-centrosymmetric quantum materials},
  journal = {Nat. Commun.},
  volume = {9},
  pages = {3740},
  year = {2018},
  doi = {10.1038/s41467-018-05759-4}
}

@article{NagaosaYanase2024,
  author = {Nagaosa, N. and Yanase, Y.},
  title = {Nonreciprocal transport and optical phenomena in quantum materials},
  journal = {Annu. Rev. Condens. Matter Phys.},
  volume = {15},
  pages = {63},
  year = {2024},
  doi = {10.1146/annurev-conmatphys-032822-033734}
}

@article{Rikken2001,
  author = {Rikken, G. L. J. A. and F{\"o}lling, J. and Wyder, P.},
  title = {Electrical magnetochiral anisotropy},
  journal = {Phys. Rev. Lett.},
  volume = {87},
  pages = {236602},
  year = {2001},
  doi = {10.1103/PhysRevLett.87.236602}
}

@article{Yokouchi2017,
  author = {Yokouchi, T. and Kanazawa, N. and Kikkawa, A. and Morikawa, D. and Shibata, K. and Arima, T. and Taguchi, Y. and Kagawa, F. and Tokura, Y.},
  title = {Electrical magnetochiral effect induced by chiral spin fluctuations},
  journal = {Nat. Commun.},
  volume = {8},
  pages = {866},
  year = {2017},
  doi = {10.1038/s41467-017-01094-2}
}

@article{Ideue2017,
  author = {Ideue, T. and Hamamoto, K. and Koshikawa, S. and Ezawa, M. and Shimizu, S. and Kaneko, Y. and Tokura, Y. and Nagaosa, N. and Iwasa, Y.},
  title = {Bulk rectification effect in a polar semiconductor},
  journal = {Nat. Phys.},
  volume = {13},
  pages = {578},
  year = {2017},
  doi = {10.1038/nphys4056}
}

@article{Li2021,
  author = {Li, Y. and Li, Y. and Li, P. and Fang, B. and Yang, X. and Wen, Y. and Zheng, D.-X. and Zhang, C.-H. and He, X. and Manchon, A. and Cheng, Z.-H. and Zhang, X.-X.},
  title = {Nonreciprocal charge transport up to room temperature in bulk {Rashba} semiconductor $\alpha$-{GeTe}},
  journal = {Nat. Commun.},
  volume = {12},
  pages = {540},
  year = {2021},
  doi = {10.1038/s41467-020-20840-7}
}

@article{Yoo2019,
  author = {Choe, D. and Jin, M.-J. and Kim, S.-I. and Choi, H.-J. and Jo, J. and Oh, I. and Park, J. and Jin, H. and Koo, H. C. and Min, B.-C. and Hong, S. and Lee, H.-W. and Baek, S.-H. and Yoo, J.-W.},
  title = {Gate-tunable giant nonreciprocal charge transport in noncentrosymmetric oxide interfaces},
  journal = {Nat. Commun.},
  volume = {10},
  pages = {4510},
  year = {2019},
  doi = {10.1038/s41467-019-12466-1}
}

@article{Wang2022,
  author = {Wang, Y. and Legg, H. F. and B{\"o}merich, T. and Park, J. and Biesenkamp, S. and Taskin, A. A. and Braden, M. and Rosch, A. and Ando, Y.},
  title = {Gigantic magnetochiral anisotropy in the topological semimetal {ZrTe$_5$}},
  journal = {Phys. Rev. Lett.},
  volume = {128},
  pages = {176602},
  year = {2022},
  doi = {10.1103/PhysRevLett.128.176602}
}

@article{Legg2022,
  author = {Legg, H. F. and R{\"o}{\ss}ler, M. and M{\"u}nning, F. and Fan, D. and Breunig, O. and Bliesener, A. and Lippertz, G. and Uday, A. and Taskin, A. A. and Loss, D. and Klinovaja, J. and Ando, Y.},
  title = {Giant magnetochiral anisotropy from quantum-confined surface states of topological insulator nanowires},
  journal = {Nat. Nanotechnol.},
  volume = {17},
  pages = {696},
  year = {2022},
  doi = {10.1038/s41565-022-01124-1}
}

@article{He2018,
  author = {He, P. and Zhang, S. S.-L. and Zhu, D. and Liu, Y. and Wang, Y. and Yu, J. and Vignale, G. and Yang, H.},
  title = {Bilinear magnetoelectric resistance as a probe of three-dimensional spin texture in topological surface states},
  journal = {Nat. Phys.},
  volume = {14},
  pages = {495},
  year = {2018},
  doi = {10.1038/s41567-017-0039-y}
}

@article{RikkenWyder2005,
  author = {Rikken, G. L. J. A. and Wyder, P.},
  title = {Magnetoelectric anisotropy in diffusive transport},
  journal = {Phys. Rev. Lett.},
  volume = {94},
  pages = {016601},
  year = {2005},
  doi = {10.1103/PhysRevLett.94.016601}
}

@article{Morimoto2016,
  author = {Morimoto, T. and Nagaosa, N.},
  title = {Chiral anomaly and giant magnetochiral anisotropy in noncentrosymmetric {Weyl} semimetals},
  journal = {Phys. Rev. Lett.},
  volume = {117},
  pages = {146603},
  year = {2016},
  doi = {10.1103/PhysRevLett.117.146603}
}

@article{Dantas2023,
  author = {Dantas, R. M. A. and Legg, H. F. and Bosco, S. and Loss, D. and Klinovaja, J.},
  title = {Determination of spin-orbit interaction in semiconductor nanostructures via nonlinear transport},
  journal = {Phys. Rev. B},
  volume = {107},
  pages = {L241202},
  year = {2023},
  doi = {10.1103/PhysRevB.107.L241202}
}

@article{zou2024nonreciprocal,
  author = {Zou, M. H. and Geng, H. and Ma, R. and Chen, W. and Sheng, L. and Xing, D. Y.},
  title = {Nonreciprocal ballistic transport in asymmetric bands},
  journal = {Phys. Rev. B},
  volume = {109},
  pages = {155302},
  year = {2024},
  doi = {10.1103/PhysRevB.109.155302}
}

@article{Leturcq2006,
  author = {Leturcq, R. and S{\'a}nchez, D. and G{\"o}tz, G. and Ihn, T. and Ensslin, K. and Driscoll, D. C. and Gossard, A. C.},
  title = {Magnetic field symmetry and phase rigidity of the nonlinear conductance in a ring},
  journal = {Phys. Rev. Lett.},
  volume = {96},
  pages = {126801},
  year = {2006},
  doi = {10.1103/PhysRevLett.96.126801}
}

@article{Angers2007,
  author = {Angers, L. and Zakka-Bajjani, E. and Deblock, R. and Gu{\'e}ron, S. and Bouchiat, H. and Cavanna, A. and Gennser, U. and Polianski, M.},
  title = {Magnetic-field asymmetry of mesoscopic dc rectification in {Aharonov}--{Bohm} rings},
  journal = {Phys. Rev. B},
  volume = {75},
  pages = {115309},
  year = {2007},
  doi = {10.1103/PhysRevB.75.115309}
}

@article{HernandezLewenkopf2009,
  author = {Hern{\'a}ndez, A. R. and Lewenkopf, C. H.},
  title = {Nonlinear conductance in a ballistic {Aharonov}--{Bohm} ring},
  journal = {Phys. Rev. Lett.},
  volume = {103},
  pages = {166801},
  year = {2009},
  doi = {10.1103/PhysRevLett.103.166801}
}

@article{vanWeperen2013,
  author = {van Weperen, I. and Plissard, S. R. and Bakkers, E. P. A. M. and Frolov, S. M. and Kouwenhoven, L. P.},
  title = {Quantized conductance in an {InSb} nanowire},
  journal = {Nano Lett.},
  volume = {13},
  pages = {387},
  year = {2013},
  doi = {10.1021/nl3035256}
}

@article{Heedt2016,
  author = {Heedt, S. and Prost, W. and Schubert, J. and Gr{\"u}tzmacher, D. and Sch{\"a}pers, Th.},
  title = {Ballistic transport and exchange interaction in {InAs} nanowire quantum point contacts},
  journal = {Nano Lett.},
  volume = {16},
  pages = {3116},
  year = {2016},
  doi = {10.1021/acs.nanolett.6b00414}
}

@article{Kammhuber2016,
  author = {Kammhuber, J. and Cassidy, M. C. and Zhang, H. and G{\"u}l, {\"O}. and Pei, F. and de Moor, M. W. A. and Nijholt, B. and Watanabe, K. and Taniguchi, T. and Car, D. and Plissard, S. R. and Bakkers, E. P. A. M. and Kouwenhoven, L. P.},
  title = {Conductance quantization at zero magnetic field in {InSb} nanowires},
  journal = {Nano Lett.},
  volume = {16},
  pages = {3482},
  year = {2016},
  doi = {10.1021/acs.nanolett.6b00051}
}

@article{Li2016,
  author = {Li, S. and Kang, N. and Fan, D. X. and Wang, L. B. and Huang, Y. Q. and Caroff, P. and Xu, H. Q.},
  title = {Coherent charge transport in ballistic {InSb} nanowire {Josephson} junctions},
  journal = {Sci. Rep.},
  volume = {6},
  pages = {24822},
  year = {2016},
  doi = {10.1038/srep24822}
}

@article{Kammhuber2017,
  author = {Kammhuber, J. and Cassidy, M. C. and Pei, F. and Nowak, M. P. and Vuik, A. and G{\"u}l, {\"O}. and Car, D. and Plissard, S. R. and Bakkers, E. P. A. M. and Wimmer, M. and Kouwenhoven, L. P.},
  title = {Conductance through a helical state in an {Indium} antimonide nanowire},
  journal = {Nat. Commun.},
  volume = {8},
  pages = {478},
  year = {2017},
  doi = {10.1038/s41467-017-00315-y}
}

@article{Peng2010,
  author = {Peng, H. and Lai, K. and Kong, D. and Meister, S. and Chen, Y. and Qi, X.-L. and Zhang, S.-C. and Shen, Z.-X. and Cui, Y.},
  title = {{Aharonov}--{Bohm} interference in topological insulator nanoribbons},
  journal = {Nat. Mater.},
  volume = {9},
  pages = {225},
  year = {2010},
  doi = {10.1038/nmat2609}
}

@article{Dufouleur2013,
  author = {Dufouleur, J. and Veyrat, L. and Teichgr{\"a}ber, A. and Neuhaus, S. and Nowka, C. and Hampel, S. and Cayssol, J. and Schumann, J. and Eichler, B. and Schmidt, O. G. and B{\"u}chner, B. and Giraud, R.},
  title = {Quasiballistic transport of {Dirac} fermions in a {Bi$_2$Se$_3$} nanowire},
  journal = {Phys. Rev. Lett.},
  volume = {110},
  pages = {186806},
  year = {2013},
  doi = {10.1103/PhysRevLett.110.186806}
}

@article{Cho2015,
  author = {Cho, S. and Dellabetta, B. and Zhong, R. and Schneeloch, J. and Liu, T. and Gu, G. and Gilbert, M. J. and Mason, N.},
  title = {{Aharonov}--{Bohm} oscillations in a quasi-ballistic three-dimensional topological insulator nanowire},
  journal = {Nat. Commun.},
  volume = {6},
  pages = {7634},
  year = {2015},
  doi = {10.1038/ncomms8634}
}

@article{Rossler2023,
  author = {R{\"o}{\ss}ler, M. and Fan, D. and M{\"u}nning, F. and Legg, H. F. and Bliesener, A. and Lippertz, G. and Uday, A. and Yazdanpanah, R. and Feng, J. and Taskin, A. and Ando, Y.},
  title = {Top-down fabrication of bulk-insulating topological insulator nanowires for quantum devices},
  journal = {Nano Lett.},
  volume = {23},
  pages = {2846},
  year = {2023},
  doi = {10.1021/acs.nanolett.3c00169}
}

@article{Manchon2015,
  author = {Manchon, A. and Koo, H. C. and Nitta, J. and Frolov, S. M. and Duine, R. A.},
  title = {New perspectives for {Rashba} spin--orbit coupling},
  journal = {Nat. Mater.},
  volume = {14},
  pages = {871},
  year = {2015},
  doi = {10.1038/nmat4360}
}

@article{vanWeperen2015,
  author = {van Weperen, I. and Tarasinski, B. and Eeltink, D. and Pribiag, V. S. and Plissard, S. R. and Bakkers, E. P. A. M. and Kouwenhoven, L. P. and Wimmer, M.},
  title = {Spin-orbit interaction in {InSb} nanowires},
  journal = {Phys. Rev. B},
  volume = {91},
  pages = {201413(R)},
  year = {2015},
  doi = {10.1103/PhysRevB.91.201413}
}

@article{angehrn2024relations,
  author = {Angehrn, G. and Legg, H. F. and Loss, D. and Klinovaja, J.},
  title = {Relations between normal state nonreciprocal transport and the superconducting diode effect in the trivial and topological phases},
  journal = {Appl. Phys. Lett.},
  volume = {125},
  pages = {023507},
  year = {2024},
  doi = {10.1063/5.0216679}
}

@article{DattaDas1990,
  author = {Datta, S. and Das, B.},
  title = {Electronic analog of the electro-optic modulator},
  journal = {Appl. Phys. Lett.},
  volume = {56},
  pages = {665},
  year = {1990},
  doi = {10.1063/1.102730}
}

@article{NadjPerge2010,
  author = {Nadj-Perge, S. and Frolov, S. M. and Bakkers, E. P. A. M. and Kouwenhoven, L. P.},
  title = {Spin--orbit qubit in a semiconductor nanowire},
  journal = {Nature},
  volume = {468},
  pages = {1084},
  year = {2010},
  doi = {10.1038/nature09682}
}

@article{Oreg2010,
  author = {Oreg, Y. and Refael, G. and von Oppen, F.},
  title = {Helical liquids and {Majorana} bound states in quantum wires},
  journal = {Phys. Rev. Lett.},
  volume = {105},
  pages = {177002},
  year = {2010},
  doi = {10.1103/PhysRevLett.105.177002}
}

@article{Lutchyn2010,
  author = {Lutchyn, R. M. and Sau, J. D. and Das Sarma, S.},
  title = {{Majorana} fermions and a topological phase transition in semiconductor--superconductor heterostructures},
  journal = {Phys. Rev. Lett.},
  volume = {105},
  pages = {077001},
  year = {2010},
  doi = {10.1103/PhysRevLett.105.077001}
}

@article{Mourik2012,
  author = {Mourik, V. and Zuo, K. and Frolov, S. M. and Plissard, S. R. and Bakkers, E. P. A. M. and Kouwenhoven, L. P.},
  title = {Signatures of {Majorana} fermions in hybrid superconductor--semiconductor nanowire devices},
  journal = {Science},
  volume = {336},
  pages = {1003},
  year = {2012},
  doi = {10.1126/science.1222360}
}

@article{Krogstrup2015,
  author = {Krogstrup, P. and Ziino, N. L. B. and Chang, W. and Albrecht, S. M. and Madsen, M. H. and Johnson, E. and Nyg{\aa}rd, J. and Marcus, C. M. and Jespersen, T. S.},
  title = {Epitaxy of semiconductor--superconductor nanowires},
  journal = {Nat. Mater.},
  volume = {14},
  pages = {400},
  year = {2015},
  doi = {10.1038/nmat4176}
}

@article{Albrecht2016,
  author = {Albrecht, S. M. and Higginbotham, A. P. and Madsen, M. and Kuemmeth, F. and Jespersen, T. S. and Nyg{\aa}rd, J. and Krogstrup, P. and Marcus, C. M.},
  title = {Exponential protection of zero modes in {Majorana} islands},
  journal = {Nature},
  volume = {531},
  pages = {206},
  year = {2016},
  doi = {10.1038/nature17162}
}

@article{Gul2018,
  author = {G{\"u}l, {\"O}. and Zhang, H. and Bommer, J. D. S. and de Moor, M. W. A. and Car, D. and Plissard, S. R. and Bakkers, E. P. A. M. and Geresdi, A. and Watanabe, K. and Taniguchi, T. and Kouwenhoven, L. P.},
  title = {Ballistic {Majorana} nanowire devices},
  journal = {Nat. Nanotechnol.},
  volume = {13},
  pages = {192},
  year = {2018},
  doi = {10.1038/s41565-017-0032-8}
}

@article{Alicea2012,
  author = {Alicea, J.},
  title = {New directions in the pursuit of {Majorana} fermions in solid state systems},
  journal = {Rep. Prog. Phys.},
  volume = {75},
  pages = {076501},
  year = {2012},
  doi = {10.1088/0034-4885/75/7/076501}
}

@article{Lutchyn2018,
  author = {Lutchyn, R. M. and Bakkers, E. P. A. M. and Kouwenhoven, L. P. and Krogstrup, P. and Marcus, C. M. and Oreg, Y.},
  title = {{Majorana} zero modes in superconductor--semiconductor heterostructures},
  journal = {Nat. Rev. Mater.},
  volume = {3},
  pages = {52},
  year = {2018},
  doi = {10.1038/s41578-018-0003-1}
}

@article{Prada2020,
  author = {Prada, E. and San-Jose, P. and de Moor, M. W. A. and Geresdi, A. and Lee, E. J. H. and Klinovaja, J. and Loss, D. and Nyg{\aa}rd, J. and Aguado, R. and Kouwenhoven, L. P.},
  title = {From {Andreev} to {Majorana} bound states in hybrid superconductor--semiconductor nanowires},
  journal = {Nat. Rev. Phys.},
  volume = {2},
  pages = {575},
  year = {2020},
  doi = {10.1038/s42254-020-0228-y}
}

@article{legg2022superconducting,
  author = {Legg, H. F. and Loss, D. and Klinovaja, J.},
  title = {Superconducting diode effect due to magnetochiral anisotropy in topological insulators and {Rashba} nanowires},
  journal = {Phys. Rev. B},
  volume = {106},
  pages = {104501},
  year = {2022},
  doi = {10.1103/PhysRevB.106.104501}
}

@article{Buttiker1988,
  author = {B{\"u}ttiker, M.},
  title = {Symmetry of electrical conduction},
  journal = {IBM J. Res. Dev.},
  volume = {32},
  pages = {317},
  year = {1988},
  doi = {10.1147/rd.323.0317}
}

@article{Buttiker1993,
  author = {B{\"u}ttiker, M.},
  title = {Capacitance, admittance, and rectification properties of small conductors},
  journal = {J. Phys.: Condens. Matter},
  volume = {5},
  pages = {9361},
  year = {1993},
  doi = {10.1088/0953-8984/5/50/017}
}

@article{ChristenButtiker1996,
  author = {Christen, T. and B{\"u}ttiker, M.},
  title = {Gauge-invariant nonlinear electric transport in mesoscopic conductors},
  journal = {Europhys. Lett.},
  volume = {35},
  pages = {523},
  year = {1996},
  doi = {10.1209/epl/i1996-00145-8}
}

@article{SanchezButtiker2004,
  author = {S{\'a}nchez, D. and B{\"u}ttiker, M.},
  title = {Magnetic-field asymmetry of nonlinear mesoscopic transport},
  journal = {Phys. Rev. Lett.},
  volume = {93},
  pages = {106802},
  year = {2004},
  doi = {10.1103/PhysRevLett.93.106802}
}

@article{SpivakZyuzin2004,
  author = {Spivak, B. and Zyuzin, A.},
  title = {Signature of the electron-electron interaction in the magnetic-field dependence of nonlinear {$I$-$V$} characteristics in mesoscopic systems},
  journal = {Phys. Rev. Lett.},
  volume = {93},
  pages = {226801},
  year = {2004},
  doi = {10.1103/PhysRevLett.93.226801}
}

@article{Song1998,
  author = {Song, A. M. and Lorke, A. and Kriele, A. and Kotthaus, J. P. and Wegscheider, W. and Bichler, M.},
  title = {Nonlinear electron transport in an asymmetric microjunction: A ballistic rectifier},
  journal = {Phys. Rev. Lett.},
  volume = {80},
  pages = {3831},
  year = {1998},
  doi = {10.1103/PhysRevLett.80.3831}
}

@article{Fleischmann2002,
  author = {Fleischmann, R. and Geisel, T.},
  title = {Mesoscopic rectifiers based on ballistic transport},
  journal = {Phys. Rev. Lett.},
  volume = {89},
  pages = {016804},
  year = {2002},
  doi = {10.1103/PhysRevLett.89.016804}
}

@article{Linke1999,
  author = {Linke, H. and Humphrey, T. E. and L{\"o}fgren, A. and Sushkov, A. O. and Newbury, R. and Taylor, R. P. and Omling, P.},
  title = {Experimental tunneling ratchets},
  journal = {Science},
  volume = {286},
  pages = {2314},
  year = {1999},
  doi = {10.1126/science.286.5448.2314}
}

@article{braunecker2005rectification,
  author = {Braunecker, B. and Feldman, D. E. and Marston, J. B.},
  title = {Rectification in one-dimensional electronic systems},
  journal = {Phys. Rev. B},
  volume = {72},
  pages = {125311},
  year = {2005},
  doi = {10.1103/PhysRevB.72.125311}
}

@article{FeldmanScheidlVinokur2005,
  author = {Feldman, D. E. and Scheidl, S. and Vinokur, V. M.},
  title = {Rectification in {Luttinger} liquids},
  journal = {Phys. Rev. Lett.},
  volume = {94},
  pages = {186809},
  year = {2005},
  doi = {10.1103/PhysRevLett.94.186809}
}

@article{Landauer1957,
  author = {Landauer, R.},
  title = {Spatial variation of currents and fields due to localized scatterers in metallic conduction},
  journal = {IBM J. Res. Dev.},
  volume = {1},
  pages = {223},
  year = {1957},
  doi = {10.1147/rd.13.0223}
}

@article{Landauer1970,
  author = {Landauer, R.},
  title = {Electrical resistance of disordered one-dimensional lattices},
  journal = {Philos. Mag.},
  volume = {21},
  pages = {863},
  year = {1970},
  doi = {10.1080/14786437008238472}
}

@article{Buttiker1986,
  author = {B{\"u}ttiker, M.},
  title = {Four-terminal phase-coherent conductance},
  journal = {Phys. Rev. Lett.},
  volume = {57},
  pages = {1761},
  year = {1986},
  doi = {10.1103/PhysRevLett.57.1761}
}

@article{beenakker1997random,
  author = {Beenakker, C. W. J.},
  title = {Random-matrix theory of quantum transport},
  journal = {Rev. Mod. Phys.},
  volume = {69},
  pages = {731},
  year = {1997},
  doi = {10.1103/RevModPhys.69.731}
}

@article{Fano1961,
  author = {Fano, U.},
  title = {Effects of configuration interaction on intensities and phase shifts},
  journal = {Phys. Rev.},
  volume = {124},
  pages = {1866},
  year = {1961},
  doi = {10.1103/PhysRev.124.1866}
}

@article{Bagwell1990,
  author = {Bagwell, P. F.},
  title = {Evanescent modes and scattering in quasi-one-dimensional wires},
  journal = {Phys. Rev. B},
  volume = {41},
  pages = {10354},
  year = {1990},
  doi = {10.1103/PhysRevB.41.10354}
}

@article{GurvitzLevinson1993,
  author = {Gurvitz, S. A. and Levinson, Y. B.},
  title = {Resonant reflection and transmission in a conducting channel with a single impurity},
  journal = {Phys. Rev. B},
  volume = {47},
  pages = {10578},
  year = {1993},
  doi = {10.1103/PhysRevB.47.10578}
}

@article{Miroshnichenko2010,
  author = {Miroshnichenko, A. E. and Flach, S. and Kivshar, Y. S.},
  title = {{Fano} resonances in nanoscale structures},
  journal = {Rev. Mod. Phys.},
  volume = {82},
  pages = {2257},
  year = {2010},
  doi = {10.1103/RevModPhys.82.2257}
}

@article{NockelStone1994,
  author = {N{\"o}ckel, J. U. and Stone, A. D.},
  title = {Resonance line shapes in quasi-one-dimensional scattering},
  journal = {Phys. Rev. B},
  volume = {50},
  pages = {17415},
  year = {1994},
  doi = {10.1103/PhysRevB.50.17415}
}

@article{Lee1999,
  author = {Lee, H.-W.},
  title = {Generic transmission zeros and in-phase resonances in time-reversal symmetric single channel transport},
  journal = {Phys. Rev. Lett.},
  volume = {82},
  pages = {2358},
  year = {1999},
  doi = {10.1103/PhysRevLett.82.2358}
}

@article{Nakatsukasa2018,
  author = {Nakatsukasa, Y. and S{\`e}te, O. and Trefethen, L. N.},
  title = {The {AAA} algorithm for rational approximation},
  journal = {SIAM J. Sci. Comput.},
  volume = {40},
  pages = {A1494},
  year = {2018},
  doi = {10.1137/16M1106122}
}

@article{Nilsson2009,
  author = {Nilsson, H. A. and Caroff, P. and Thelander, C. and Larsson, M. and Wagner, J. B. and Wernersson, L.-E. and Samuelson, L. and Xu, H. Q.},
  title = {Giant, level-dependent $g$ factors in {InSb} nanowire quantum dots},
  journal = {Nano Lett.},
  volume = {9},
  pages = {3151},
  year = {2009},
  doi = {10.1021/nl901333a}
}

@misc{SM,
  note = {See Supplemental Material at the end of this article, which also cites Refs.~\citenum{Nakatsukasa2018,Nilsson2009}, for the symmetry proofs, gate electrostatics, Born expansions, T-matrix calculations, resonant devices, and all parameters}
}

@misc{Zenodo,
  author = {Osborne, T. J. and Braunecker, B. and Legg, H. F.},
  title = {Data and code for ``{Quantum} {Interference} {Enhanced} {Rectification} in {Mesoscopic} {Nanowires}''},
  howpublished = {Zenodo, doi:10.5281/zenodo.23529262},
  year = {2026},
  doi = {10.5281/zenodo.23529262}
}

\end{document}


\title{Supplemental Material for\\ ``Quantum Interference Enhanced Rectification in Mesoscopic Nanowires''}
\author{T. J. Osborne}
\author{B. Braunecker}
\author{H. F. Legg}
\affiliation{School of Physics and Astronomy, University of St Andrews, North Haugh, St Andrews KY16 9SS, United Kingdom}
\maketitle

In this Supplemental Material (SM), we prove the symmetry constraints on two-terminal rectification used in the main text (Sec.~\ref{sec:symmetry}) and derive the bias dependence of the gate potentials (Sec.~\ref{sec:electrostatics}). We then give the formal solution of the scattering problem, from which both the exact transmission and the scattering states follow (Sec.~\ref{sec:tmatrix}). Expanding this solution gives the Born series of the 1D and quasi-1D models (Secs.~\ref{sec:1d} and~\ref{sec:q1d}). We also analyze the resonant quasi-1D device of Fig.~3 of the main text and its 1D counterpart (Sec.~\ref{sec:resonant}). Finally, we list all parameters (Sec.~\ref{sec:parameters}) and compare the resonant three-gate device with a single gate (Sec.~\ref{sec:single}). Throughout, equation and figure numbers without the prefix S refer to the main text.

\section{Symmetry constraints on rectification}
\label{sec:symmetry}

At a fixed energy, let $a_L,a_R$ ($b_L,b_R$) collect the incoming (outgoing) unit-flux channel amplitudes in the left and right leads. The scattering matrix of Eq.~(3) is defined by
\begin{equation}
 \begin{pmatrix}b_L\\ b_R\end{pmatrix}
 =S(E,V)\begin{pmatrix}a_L\\ a_R\end{pmatrix},\qquad
 S=\begin{pmatrix}r&t'\\t&r'\end{pmatrix},
 \label{eq:S-def}
\end{equation}
such that $\mathcal T=\Tr(t^\dagger t)$ and $\mathcal T'=\Tr(t'^\dagger t')$. Let $H_{\rm leads}(V)$ describe the leads decoupled from the device and $H_{\rm tot}(V)$ the coupled system. The M\o ller operators are
\begin{equation}
 \Omega_{\rm in/out}(V)=\operatorname*{s-lim}_{t\to\mp\infty}
 e^{iH_{\rm tot}(V)t/\hbar}e^{-iH_{\rm leads}(V)t/\hbar},
 \qquad S(V)=\Omega_{\rm out}^\dagger(V)\Omega_{\rm in}(V),
 \label{eq:moller}
\end{equation}
where we assume their existence and asymptotic completeness on the propagating subspace.

We consider transformations that map the leads onto the leads, such as reflections and spin rotations, and call the restriction of such a transformation to the decoupled leads its action on the leads. Suppose first that a unitary transformation $\mathcal U$, with action $u$ on the leads, obeys
\begin{equation}
 H_{\rm tot}(-V)=\mathcal U H_{\rm tot}(V)\mathcal U^{-1},\qquad
 H_{\rm leads}(-V)=uH_{\rm leads}(V)u^{-1}.
 \label{eq:unitary-H-map}
\end{equation}
Inserting this into Eq.~\eqref{eq:moller} gives
\begin{equation}
 \Omega_{\rm in/out}(-V)=\mathcal U\Omega_{\rm in/out}(V)u^{-1},\qquad
 S(-V)=uS(V)u^{-1}.
 \label{eq:unitary-S-map}
\end{equation}
If $u$ exchanges the leads, $uP_Lu^{-1}=P_R$, where $P_{L/R}$ project onto the channels of the left (right) lead, its restrictions $u_R=P_RuP_L$ and $u_L=P_LuP_R$ are unitary maps between the lead-channel spaces. Therefore
\begin{equation}
 t(-V)=u_Rt'(V)u_L^{-1},\qquad
 \mathcal T(-V)=\mathcal T'(V).
 \label{eq:unitary-t-map}
\end{equation}
Since $S^\dagger S=SS^\dagger=1$ for an elastic two-terminal problem, comparing the traces of the left-lead blocks gives the standard identity~\cite{Buttiker1988}
\begin{equation}
 \mathcal T(E,V)=\mathcal T'(E,V).
 \label{eq:two-terminal-T}
\end{equation}
Together, Eqs.~\eqref{eq:unitary-t-map} and~\eqref{eq:two-terminal-T} establish $\mathcal T(E,-V)=\mathcal T(E,V)$. If $u$ does not exchange the leads, Eq.~\eqref{eq:unitary-S-map} gives $t(-V)=u_Rt(V)u_L^{-1}$ directly.

For an antiunitary transformation $\mathcal A$ with action $w$ on the leads, which obeys Eq.~\eqref{eq:unitary-H-map} with $\mathcal U\to\mathcal A$ and $u\to w$, $\mathcal A e^{iH(V)t/\hbar}\mathcal A^{-1}=e^{-iH(-V)t/\hbar}$ for $H=H_{\rm tot},H_{\rm leads}$. Consequently, the incoming and outgoing M\o ller operators are exchanged,
\begin{equation}
 \Omega_{\rm in}(-V)=\mathcal A\Omega_{\rm out}(V)w^{-1},\qquad
 \Omega_{\rm out}(-V)=\mathcal A\Omega_{\rm in}(V)w^{-1},
 \label{eq:anti-moller}
\end{equation}
such that
\begin{equation}
 S(-V)=w S^\dagger(V)w^{-1}.
 \label{eq:anti-S}
\end{equation}
If $w$ exchanges the leads, writing its restrictions $w_R=P_RwP_L$ and $w_L=P_LwP_R$ as $w_{R/L}=W_{R/L}\mathcal K$, where $W_{R/L}$ are unitary matrices and $\mathcal K$ is complex conjugation, gives
\begin{equation}
 t(-V)=W_Rt^{T}(V)W_L^\dagger.
 \label{eq:anti-t}
\end{equation}
The Frobenius norm is unchanged by transposition and unitary rotations, so that again $\mathcal T(E,-V)=\mathcal T(E,V)$. If $w$ does not exchange the leads, $t'$ replaces $t$ on the right-hand side of Eq.~\eqref{eq:anti-t}, and Eq.~\eqref{eq:two-terminal-T} completes the proof. This proves Eq.~(4) of the main text for both classes of transformation. We note that the absence of such a map is necessary, but not sufficient, for a finite rectified current.

\section{Electrostatic origin of the bias-dependent gate potentials}
\label{sec:electrostatics}

We consider metallic finger gates that are electrically floating. Let $\Phi_j$ and $Q_j$ be the electrostatic potential and fixed charge of gate $j$. With capacitances $C_{jL}$ and $C_{jR}$ to the two reservoirs and a total capacitance $C_{j0}$ to conductors with potentials $V_0$ that remain fixed as the source--drain bias is swept, charge conservation gives
\begin{equation}
 Q_j=C_{jL}(\Phi_j-V_L)+C_{jR}(\Phi_j-V_R)
 +C_{j0}(\Phi_j-V_0).
 \label{eq:charge-gate}
\end{equation}
For $V_L=-V_R=V/2$ and $C_{j\Sigma}=C_{jL}+C_{jR}+C_{j0}$, the gate potential is
\begin{equation}
 \Phi_j(V)=\Phi_j(0)+\frac{C_{jL}-C_{jR}}{2C_{j\Sigma}}V.
 \label{eq:phi-j}
\end{equation}
In other words, a floating gate follows the potential of the lead to which it is more strongly coupled. A gate at the center of a left--right-symmetric device has $C_{jL}=C_{jR}$ and is therefore bias independent, as is the central gate $U_c$ of the main text. For the two outer gates, mirror symmetry gives
\begin{equation}
 C_{LL}=C_{RR}=C_>,\qquad C_{LR}=C_{RL}=C_<,
 \qquad C_{L\Sigma}=C_{R\Sigma}=C_\Sigma,
 \label{eq:cap-mirror}
\end{equation}
such that $\Phi_L(V)=\Phi_L(0)+\zeta V$ and $\Phi_R(V)=\Phi_R(0)-\zeta V$, with $\zeta=(C_>-C_<)/(2C_\Sigma)$.

To relate the gate potentials to the potential energy in the wire, let $u_j(\bm r)$ be the characteristic potential~\cite{ChristenButtiker1996} that equals unity on gate $j$ and zero on the other metallic boundaries. In the linear electrostatic regime it obeys
\begin{equation}
 \bm\nabla\cdot[\epsilon(\bm r)\bm\nabla u_j(\bm r)]=0,
 \qquad
 \delta\varphi(\bm r;V)=\zeta V[u_L(\bm r)-u_R(\bm r)],
 \label{eq:characteristic-potential}
\end{equation}
where $\epsilon(\bm r)$ is the dielectric function and $\delta\varphi$ the bias-induced potential. We neglect the potential drop along the wire itself, described by the characteristic potentials of the leads, and retain only the part induced by the gates. Mirror symmetry of the dielectric and conductor geometry gives $u_L(-x,\bm r_\perp)=u_R(x,\bm r_\perp)$ and hence $\delta\varphi(-x,\bm r_\perp;V)=\delta\varphi(x,\bm r_\perp;-V)$. In particular, $\delta\varphi$ vanishes at the center of the device, $x=0$, as does the neglected potential drop along the wire for the symmetric bias $V_L=-V_R$, such that the potential at the center remains at its equilibrium value for both polarities. Projecting the induced potential energy onto the transverse modes and integrating over each short gate profile gives the point-like gate strengths of Eq.~(5),
\begin{equation}
 U_{L/R}(V)=U_0\pm\Gamma eV,
 \qquad U(x,-V)=U(-x,V),
 \label{eq:local-U}
\end{equation}
where $\Gamma$ absorbs $\zeta$ and the overlap of $\delta\varphi$ with the transverse modes and has units of length.

\section{Scattering formalism}
\label{sec:tmatrix}

Both models of the main text describe a clean wire with Hamiltonian $H(k)$, Eq.~(6) or~(9), and point-like scatterers, $U(x,V)=\sum_jU_j(V)\delta(x-x_j)$, with matrix strengths $U_j$ that act in spin (and subband) space. All results below, the Born expansions of Secs.~\ref{sec:1d} and~\ref{sec:q1d} as well as the exact transmission, follow from one formal solution of this scattering problem, which we give first.

\subsection{T matrix and reflection amplitudes}

The clean retarded Green function is
\begin{equation}
 g^R(x;E)=\int\frac{dk}{2\pi}\,e^{ikx}\left[E+i0^+-H(k)\right]^{-1},
 \label{eq:green}
\end{equation}
which we denote by $g_1^R$ for the 1D Hamiltonian $H_1$ and by $g_2^R$ for the quasi-1D Hamiltonian $H_2$. For an incident wave $\psi_{\rm in}(x)$ of the clean wire at energy $E$, the scattering state obeys the Lippmann--Schwinger equation, which for point-like scatterers reads
\begin{equation}
 \psi(x)=\psi_{\rm in}(x)+\sum_jg^R(x-x_j;E)\,U_j\,\psi(x_j).
 \label{eq:lippmann-schwinger}
\end{equation}
Evaluating Eq.~\eqref{eq:lippmann-schwinger} at the scatterers gives a finite set of linear equations. With the block matrices
\begin{equation}
 \mathbf G_{jl}(E)=g^R(x_j-x_l;E),\qquad
 \mathbf U(V)=\operatorname{diag}(U_j),
 \label{eq:contact-matrices}
\end{equation}
its solution is $\psi(x)=\psi_{\rm in}(x)+\sum_{jl}g^R(x-x_j;E)\,\mathbf T_{jl}(E,V)\,\psi_{\rm in}(x_l)$, with the exact finite-rank T matrix~\cite{Bagwell1990}
\begin{equation}
 \mathbf T(E,V)=\left[\mathbf 1-\mathbf U(V)\mathbf G(E)\right]^{-1}\mathbf U(V)
 =\mathbf U+\mathbf U\mathbf G\mathbf U+\mathbf U\mathbf G\mathbf U\mathbf G\mathbf U+\cdots.
 \label{eq:exact-T}
\end{equation}
Far to the left of the scatterers, only the propagating poles of $g^R$ survive. If $\alpha$ and $\beta$ denote incoming and reflected propagating modes with wavevectors $k_\alpha$ and $k_\beta$, spinors $\bm u_\alpha$ and $\bm u_\beta$, and velocities $v_\alpha$ and $v_\beta$, the flux-normalized reflection amplitudes are therefore
\begin{equation}
 r_{\beta\alpha}(E,V)=-\frac{i}{\hbar\sqrt{|v_\beta v_\alpha|}}
 \sum_{jl}e^{-ik_\beta x_j}\bm u_\beta^\dagger
 \mathbf T_{jl}(E,V)e^{ik_\alpha x_l}\bm u_\alpha.
 \label{eq:reflection-T}
\end{equation}
For a lossless wire, $\mathcal T(E,V)=N_{\rm ch}(E)-\sum_{\alpha,\beta}|r_{\beta\alpha}|^2$, where $N_{\rm ch}(E)$ is the number of propagating channels at energy $E$. We verify unitarity, $|r|^2+|t|^2=1$, to $10^{-12}$ in the single-channel regime.

In the single-channel regime, with the open channel at wavevector $\pm k_E$, spinor $\bm u(\pm k_E)$, and velocity $\pm v_E$, Eq.~\eqref{eq:reflection-T} for incidence from the left becomes
\begin{equation}
 r=-\frac{ib}{\hbar|v_E|},\qquad
 b=\sum_{jl}e^{ik_E(x_j+x_l)}\,\bm u^\dagger(-k_E)\,\mathbf T_{jl}\,\bm u(k_E),\qquad
 \mathcal T=1-|r|^2.
 \label{eq:reduced-amplitude}
\end{equation}
The Born expansion is the expansion of $\mathbf T$ in Eq.~\eqref{eq:exact-T} in powers of the scatterer strengths. The $n$th term of this expansion, with $n$ factors of $\mathbf U$ and $n-1$ factors of $\mathbf G$, gives the reduced amplitude $b^{(n)}=O(U^n)$ that we use in Secs.~\ref{sec:1d} and~\ref{sec:q1d}. For incidence from the right, $k_E\to-k_E$ in Eq.~\eqref{eq:reduced-amplitude}.

\subsection{Numerical evaluation}

For the exact results, we evaluate Eq.~\eqref{eq:exact-T} without expansion and obtain the clean Green functions directly in the continuum by summing the residues of the poles of $[E+i0^+-H(k)]^{-1}$ in the appropriate half-plane, which includes all evanescent modes. Since $f_L-f_R$ vanishes outside the bias window, we obtain the currents by integrating Eq.~(1) over $-e|V|/2-14k_{\mathrm B}T\leq E\leq e|V|/2+14k_{\mathrm B}T$ with the trapezoidal rule on 81 to 4001 energies, depending on the calculation. Changing the grid between 81 and 1601 energies changes $\Delta G/G_0$ by less than $10^{-4}$.

\subsection{Scattering states}

For an electron that arrives from the left in the open channel $\alpha$, the incident wave is the flux-normalized plane wave $\psi_{\rm in}(x)=e^{ik_\alpha x}\bm u_\alpha/\sqrt{\hbar|v_\alpha|}$. The scattering state shown in Fig.~3(b) is the solution of Eq.~\eqref{eq:lippmann-schwinger} given above,
\begin{equation}
 \psi(x)=\psi_{\rm in}(x)+\sum_{jl}g^R(x-x_j;E)\,\mathbf T_{jl}(E,V)\,\psi_{\rm in}(x_l).
 \label{eq:scattering-state}
\end{equation}
We split the four components of $\psi(x)$ into the two subbands, $|\psi_\tau(x)|^2=\sum_\sigma|\psi_{\sigma\tau}(x)|^2$ with $\tau=1,2$, and divide by the uniform density $|\psi_{\rm in}|^2=1/\hbar|v_\alpha|$ of the incident wave. This ratio is independent of the normalization of $\psi_{\rm in}$ and equals one in the clean wire. On the gates it exceeds one near a resonance, since the electron dwells there, whereas to the left of the gates the interference of the incident and reflected waves gives values between zero and four for full reflection.

\section{1D model}
\label{sec:1d}

\subsection{Spectrum, symmetry, and overlaps}

For the Hamiltonian of Eq.~(6), the two clean bands are
\begin{equation}
 E_\pm(k)=\frac{\hbar^2k^2}{2m^*}-\mu
 \pm D(k),\qquad D(k)=\sqrt{(\alpha_Rk)^2+\Delta_x^2},
 \label{eq:one-spectrum}
\end{equation}
and a convenient lower-band spinor is
\begin{equation}
 u_-(k)=\frac{1}{\sqrt2}
 \begin{pmatrix}-e^{-i\vartheta(k)}\\1\end{pmatrix},\qquad
 \sin\vartheta(k)=\frac{\alpha_Rk}{D(k)},\quad
 \cos\vartheta(k)=\frac{\Delta_x}{D(k)}.
 \label{eq:spinor-one}
\end{equation}
Since $\sigma_xH_1^*(k)\sigma_x=H_1(k)$, the clean wire is invariant under the antiunitary map $\mathcal P\sigma_x\mathcal K$, where $\mathcal P$ takes $x\to-x$ and $\mathcal K$ is complex conjugation. Combined with bias reversal, this map leaves two scalar gates with the mirror-related strengths of Eq.~\eqref{eq:local-U} invariant and exchanges the leads, such that it enforces reciprocity (Sec.~\ref{sec:symmetry}). Since $\sigma_x\sigma_z^*\sigma_x=-\sigma_z$, the magnetic scatterer $M\sigma_z$ at the origin is odd under this map and therefore removes the constraint. The field need not point along the wire. For a single subband, a spin rotation about $y$, which leaves $\alpha_Rk\sigma_y$ and the scalar gates unchanged, maps a field anywhere in the $xz$ plane onto one along $x$ with the same Zeeman energy and rotates the moment of the magnetic scatterer by the same angle. Only the angle between the field and this moment therefore matters. A moment parallel to the field gives no rectification, since it is even under $\mathcal P\sigma_x\mathcal K$. The complete potential of the 1D device is
\begin{equation}
 U_{\rm 1D}(x,V)=U_L(V)\delta(x+d/2)\sigma_0
 +M\delta(x)\sigma_z
 +U_R(V)\delta(x-d/2)\sigma_0.
 \label{eq:one-contact-potential}
\end{equation}

At energy $E$, let $k_E>0$ solve $E_-(k_E)=E$ and let $v_E\equiv v(k_E)=\hbar^{-1}\partial_kE_-(k)|_{k_E}$ be the group velocity of the main text. For an electron incident from the left, the first-order reflection from $+k_E$ to $-k_E$ involves the overlaps
\begin{equation}
 m_s=u_-^\dagger(-k_E)u_-(k_E)=e^{-i\vartheta_E}\cos\vartheta_E,\qquad
 m_M=u_-^\dagger(-k_E)\sigma_z u_-(k_E)=-i\tan\vartheta_E\,m_s,
 \label{eq:one-overlaps}
\end{equation}
where $\vartheta_E=\vartheta(k_E)$, such that $\tan\vartheta_E=\alpha_Rk_E/\Delta_x$. The magnetic scatterer therefore reflects with a phase shifted by $\pi/2$ relative to a scalar gate, with the sign set by that of $M\alpha_Rk_E/\Delta_x$, i.e., by $+\pi/2$ for $M<0$ and $\Delta_x>0$, as drawn in Fig.~2(a).

\subsection{Born expansion}

With $\bm u=u_-$ in Eq.~\eqref{eq:reduced-amplitude}, the first-order term $\mathbf T\approx\mathbf U$ of Eq.~\eqref{eq:exact-T} gives the reduced amplitude
\begin{equation}
 b^{(1)}=m_s\left[U_L(V)e^{-ik_Ed}+U_R(V)e^{ik_Ed}\right]+Mm_M
 =m_s\left[2U_0\cos(k_Ed)-2i\Gamma eV\sin(k_Ed)-iM\tan\vartheta_E\right],
 \label{eq:one-b1}
\end{equation}
where each gate at $x_j$ contributes the propagation phase $e^{2ik_Ex_j}$. Since $\mathcal T=1-|r|^2$, the bias-odd part at second order is
\begin{equation}
 \mathcal T_{\rm odd}^{(2)}(E,V)=
 -\frac{4M\Gamma eV}{\hbar^2|v_E|^2}
 \frac{\alpha_Rk_E\Delta_x}{(\alpha_Rk_E)^2+\Delta_x^2}
 \sin(k_Ed),
 \label{eq:one-P-odd}
\end{equation}
which is Eq.~(7) of the main text, since $|m_s|^2\tan\vartheta_E=\alpha_Rk_E\Delta_x/D_E^2$ with $D_E=D(k_E)$. For the next order, the term $\mathbf U\mathbf G\mathbf U$ of Eq.~\eqref{eq:exact-T}, with the Green function $g_1^R$ of Eq.~\eqref{eq:green}, gives the second-order amplitude
\begin{equation}
 b^{(2)}=\sum_{j,l}e^{ik_E(x_j+x_l)}
 u_-^\dagger(-k_E)U_jg_1^R(x_j-x_l;E)U_lu_-(k_E),
 \label{eq:one-b2}
\end{equation}
where $(x_j,U_j)=(-d/2,U_L\sigma_0)$, $(0,M\sigma_z)$, and $(d/2,U_R\sigma_0)$. The transmission up to third order in the scatterer strengths is then
\begin{equation}
 \mathcal T^{(\leq3)}=1-\frac{1}{\hbar^2|v_E|^2}
 \bigl\{|b^{(1)}|^2+2\re[b^{(1)*}b^{(2)}]\bigr\},
 \label{eq:one-P3}
\end{equation}
and its bias-odd part gives the $O(U^3)$ curve of Fig.~2(c).

\section{Quasi-1D model}
\label{sec:q1d}

\subsection{Origin of the transverse subband coupling}

We start from the confined continuum Hamiltonian
\begin{equation}
 H=\frac{p_x^2+p_y^2+p_z^2}{2m^*}+V_\perp(y,z)-\mu
 +\frac{\alpha_R}{\hbar}(\sigma_yp_x-\sigma_xp_y)
 +\Delta_x\sigma_x+\Delta_z\sigma_z+U(x,y,z;V),
 \label{eq:3d-H}
\end{equation}
where $V_\perp$ is the confining potential. Projecting onto the two lowest real transverse wave functions $|1\rangle$ and $|2\rangle$, with energies $\epsilon_1$ and $\epsilon_2$, gives $\langle1|p_y|2\rangle=i\hbar q_{12}$ and
\begin{equation}
 \Delta_\perp=\epsilon_2-\epsilon_1,\qquad
 \lambda_\perp=\alpha_Rq_{12},
 \label{eq:transverse-definitions}
\end{equation}
which yields the Hamiltonian of Eq.~(9). Both subbands keep the effective mass $m^*$, since the confinement only adds the transverse energies $\epsilon_n$ to the motion along the wire. This holds for any confinement that does not depend on $x$, not only for hard walls. For hard-wall confinement of width $W_y$,
\begin{equation}
 \Delta_\perp=\frac{3\hbar^2\pi^2}{2m^*W_y^2},\qquad
 q_{12}=\frac{8}{3W_y},\qquad
 \lambda_\perp=\lambda_0\equiv\frac{8\alpha_R}{3W_y}.
 \label{eq:hardwall}
\end{equation}
A gate potential that is approximately uniform over the cross section is proportional to $\sigma_0\tau_0$. For $\Delta_z=0$, the clean-wire Hamiltonian $H_2(k)$ of Eq.~(9) obeys $\tau_z\sigma_xH_2^*(k)\sigma_x\tau_z=H_2(k)$, such that the map $\mathcal P\tau_z\sigma_x\mathcal K$ relates the two bias orientations, whereas for $\Delta_x=0$, $\sigma_zH_2^*(k)\sigma_z=H_2(k)$ and the corresponding map is $\mathcal P\sigma_z\mathcal K$. Transverse Rashba coupling together with $\Delta_x\Delta_z\neq0$ removes both constraints. Since the exact calculation then gives $\Delta G\neq0$ [see Figs.~2(d) and~3 of the main text], no other forward--reverse map can exist either (Sec.~\ref{sec:symmetry}).

The clean quasi-1D Hamiltonian also obeys $\tau_zH_2^*(-k)\tau_z=H_2(k)$. The clean wire therefore possesses the antiunitary symmetry $\Theta=\tau_z\mathcal K$, with $\Theta^2=1$, which reverses the momentum without exchanging the leads and leaves the real scalar gate potentials invariant at any bias, as we use in Sec.~\ref{sec:resonant}.

\subsection{Field parity and orbital effects}
\label{sec:orbital}

Time reversal $i\sigma_y\mathcal K$ maps the device at field $\bm B$ onto that at $-\bm B$ without exchanging the leads, such that $\mathcal T(E,V;\bm B)=\mathcal T(E,V;-\bm B)$, whereas the two maps above give $\mathcal T(E,V;\Delta_z)=\mathcal T(E,-V;-\Delta_z)$ and $\mathcal T(E,V;\Delta_x)=\mathcal T(E,-V;-\Delta_x)$. Hence $\Delta G$ is even in $\bm B$ but odd in $\Delta_x$ and in $\Delta_z$.

The calculations of the main text include the field only through the Zeeman energy. A field component $B_z$ also couples to the orbital motion across the nanowire through $A_x=-B_zy$. Projected onto the two subbands, this adds the subband mixing $-\hbar\omega_cy_{12}k\tau_x$, with $\omega_c=eB_z/m^*$ and $y_{12}=\langle1|y|2\rangle=-16W_y/(9\pi^2)$, and diamagnetic shifts of the subbands. These terms vanish for $B_z=0$ and preserve $\mathcal P\sigma_z\mathcal K$ and $\Theta$, such that the constraints above and the field parity are unchanged. For $|g|=50$ at the operating point of Eq.~\eqref{eq:par-two-strong}, however, $B_z\approx0.18$~T, the mixing reaches about $0.8$~meV at $k_F$, comparable to $\lambda_\perp$, and the lowest band rises by about $0.3$~meV, which detunes the resonance and suppresses $\Delta G$. Retuning all parameters with the orbital terms included gives $\Delta G/G_0\approx0.79$, with $G(+V)\approx0.94G_0$ and $G(-V)\approx0.15G_0$, for $\Delta\approx1.8$~meV, $\theta\approx78^\circ$, $\mu\approx-1.2$~meV, $U_0\approx-260$~meV\,nm, $U_c\approx-490$~meV\,nm, $d\approx73$~nm, and $e|V|\approx0.46$~meV (the code~\cite{Zenodo} gives the exact values). Since the magnetic length at this point, about $50$~nm, is below $W_y$, a quantitative treatment requires more than two subbands.

\subsection{Born expansion through fifth order}

The three scalar gates are described by
\begin{equation}
 U_{\rm q1D}(x,V)=\left[U_L(V)\delta(x+d/2)+U_c\delta(x)
 +U_R(V)\delta(x-d/2)\right]\sigma_0\tau_0.
 \label{eq:two-contact-potential}
\end{equation}
We assume that only the lowest clean band propagates in the bias window and let $k_E>0$ and $v_E$ be the wavevector and group velocity of its right-moving state at energy $E$, with $\bm u_0(k)$ the four-component eigenstate. Since $\mathcal T(E,V)=\mathcal T'(E,V)$ [Eq.~\eqref{eq:two-terminal-T}], we consider here reflection from $-k_E$ to $+k_E$, i.e., incidence from the right, which reverses the sign of the $\Gamma eV$ term in $b^{(1)}$ relative to Eq.~\eqref{eq:one-b1} but leaves $\mathcal T$ unchanged. With the Green function $g_2^R$ of Eq.~\eqref{eq:green} for $H_2$, we define
\begin{equation}
 m_E=\bm u_0^\dagger(k_E)\bm u_0(-k_E),\qquad
 \mathcal G_E(x)=\bm u_0^\dagger(k_E)g_2^R(x;E)\bm u_0(-k_E),
 \label{eq:two-green}
\end{equation}
where $\mathcal G_E$ projects the full $4\times4$ Green function, including virtual propagation through all spin and subband states, onto the counterpropagating states. The first-order reduced amplitude of Eq.~\eqref{eq:reduced-amplitude}, with $k_E\to-k_E$, is
\begin{equation}
 b^{(1)}=m_E\left[U_L(V)e^{ik_Ed}+U_c+U_R(V)e^{-ik_Ed}\right]
 =m_E\left[U_c+2U_0\cos(k_Ed)+2i\Gamma eV\sin(k_Ed)\right].
 \label{eq:two-b1}
\end{equation}
Since all three gates share the overlap $m_E$, $|b^{(1)}|^2$ is even in bias. The leading bias-odd transmission therefore comes from interference between $b^{(1)}$ and the second-order amplitude
\begin{equation}
 b^{(2)}=\sum_{j,l}e^{-ik_E(x_j+x_l)}
 \bm u_0^\dagger(k_E)U_jg_2^R(x_j-x_l;E)U_l\bm u_0(-k_E),
 \label{eq:two-b2-general}
\end{equation}
which gives
\begin{equation}
\mathcal T_{\rm odd}^{(3)}(E,V)=-\frac{4\Gamma eV}{\hbar^2|v_E|^2}\,\im\big\{m_E^*\big[A_E+U_c(U_c-2U_0)B_E\big]\big\},
 \label{eq:two-P-odd-SM}
\end{equation}
with
\begin{equation}
 \begin{aligned}
 A_E&=(U_0^2-\Gamma^2e^2V^2)\sin(k_Ed)\,[\mathcal G_E(d)+\mathcal G_E(-d)-2\cos(k_Ed)\,\mathcal G_E(0)],\\
 B_E&=\sin(k_Ed)\,\mathcal G_E(0)-\sin(k_Ed/2)\,[\mathcal G_E(d/2)+\mathcal G_E(-d/2)].
 \end{aligned}
 \label{eq:AB}
\end{equation}
The term $A_E$ arises from paths that involve the two outer gates, separated by $d$, whereas $B_E$ contains the paths through the central gate and therefore introduces propagation phases associated with both $d$ and $d/2$. For $U_c=0$, only $A_E$ remains, as in the two-gate device.

For the Born curves of Fig.~2(d), we retain every contribution to the transmission at a given total order. With $b^{(n)}=O(U^n)$, we define
\begin{equation}
 C_n(V)=\sum_{p+q=n}b^{(p)*}(V)b^{(q)}(V),\qquad
 \mathcal T^{(\leq N)}=1-
 \frac{1}{\hbar^2|v_E|^2}
 \sum_{n=2}^{N}C_n,
 \label{eq:two-consistent-born}
\end{equation}
with, in particular,
\begin{equation}
 C_2=|b^{(1)}|^2,\quad
 C_3=2\re[b^{(1)*}b^{(2)}],\quad
 C_4=|b^{(2)}|^2+2\re[b^{(1)*}b^{(3)}],\quad
 C_5=2\re[b^{(1)*}b^{(4)}+b^{(2)*}b^{(3)}].
 \label{eq:two-C2-C5}
\end{equation}
The $n$th amplitude follows from the term $\mathbf U(\mathbf G\mathbf U)^{n-1}$ of Eq.~\eqref{eq:exact-T}, with the full $4\times4$ Green function at each intermediate propagation. In Fig.~2(d), the odd contributions at $O(U^3)$ and $O(U^4)$ nearly coincide, whereas the small visible correction at $O(U^5)$ comes from $C_5$. Over the range shown, the $O(U^3)$ and $O(U^4)$ results deviate from the exact $\Delta G/G_0$ by up to about $20\%$ of its maximum, whereas the $O(U^5)$ result agrees with it to within about $3\%$.

\section{Resonant devices}
\label{sec:resonant}

In this section, we characterize the resonance of the quasi-1D three-gate device of Fig.~3 at the parameters of Eq.~\eqref{eq:par-two-strong} as well as the corresponding resonant 1D device.

\subsection{Uncoupled subbands}

For $\lambda_\perp=0$ the two subbands decouple, since the scalar gates are proportional to $\sigma_0\tau_0$ and cannot transfer an electron between them. An electron incident in the lowest subband then scatters from three strongly attractive point-like gates. We find $G(+V)=G(-V)\approx0.015G_0$, i.e., $\Delta G=0$, with the transmission at most about $0.05$ across the bias window even at $V=0$. The vanishing of $\Delta G$ is expected, since the lowest subband alone is the 1D model of Eq.~(6) with an additional $\Delta_z\sigma_z$ and scalar gates. With $\hat{\bm n}=(\sin\theta,0,\cos\theta)$ the field direction, $(\hat{\bm n}\cdot\bm\sigma)H(-k)(\hat{\bm n}\cdot\bm\sigma)=H(k)$ for this Hamiltonian, such that the lead-exchanging unitary map $\mathcal P\,\hat{\bm n}\cdot\bm\sigma$ combined with bias reversal enforces reciprocity (Sec.~\ref{sec:symmetry}).

\subsection{Bound states of the second subband}

For $\lambda_\perp=0$, the second subband obeys the same Hamiltonian as the lowest one, shifted in energy by $\Delta_\perp$, with its band edge at about $1.8$~meV. The attractive gates bind states below this edge, at the energies at which $\mathbf 1-\mathbf U\mathbf G$ of Eq.~\eqref{eq:exact-T}, restricted to the second subband, becomes singular. At $\pm V$, the outer gates bind a pair of states at $E\approx-0.28$ and $-0.10$~meV, which at $V=0$ lie at $-0.27$ and $-0.12$~meV and are shared equally by the two outer gates, whereas at finite bias each localizes predominantly on one gate. The central gate binds a further pair far below the bias window, at about $-34$ and $-32$~meV. In isolation, bias reversal therefore leaves the bound-state energies unchanged and mirrors their densities, such that rectification arises only through their coupling to the open channel.

\subsection{Dependence on the transverse coupling}

\begin{figure}[t]
    \centering
    \includegraphics{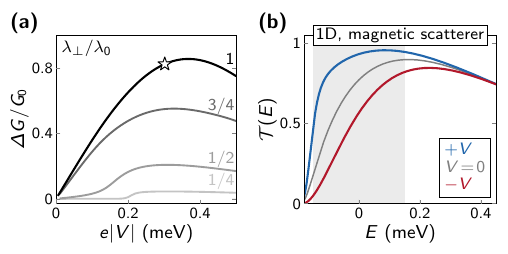}
    \caption{\textbf{Transverse coupling and the 1D resonant device.} (a) Diode effectiveness $\Delta G/G_0$ of the resonant quasi-1D device as a function of bias magnitude for the transverse coupling scaled to $\lambda_\perp/\lambda_0=1/4$, $1/2$, $3/4$, and $1$, where $\lambda_0$ is the hard-wall value of Eq.~\eqref{eq:hardwall}; $\Delta G=0$ for $\lambda_\perp=0$. The star marks the operating point of Fig.~3. Other parameters as in Eq.~\eqref{eq:par-two-strong}. (b) Transmission of the resonant 1D device with a magnetic central scatterer at $+V$ (blue), $V=0$ (gray), and $-V$ (red), with the bias window shaded, for the parameters of Eq.~\eqref{eq:par-one-strong}.}
    \label{fig:lambda}
\end{figure}

Scaling the transverse coupling as $\lambda_\perp=f\lambda_0$ gives $\Delta G/G_0\approx0.05$, $0.21$, $0.55$, and $0.82$ at the operating bias for $f=1/4$, $1/2$, $3/4$, and $1$, respectively [see Fig.~\ref{fig:lambda}(a)], with $\Delta G/G_0\approx0.7f^2$ for $0.1\lesssim f\lesssim0.3$.

\subsection{Transmission zeros}
\label{sec:zeros}

The symmetry $\Theta=\tau_z\mathcal K$ of Sec.~\ref{sec:q1d} reverses the momentum without exchanging the leads. In single-channel transport, such a symmetry ensures that transmission zeros occur generically on the real energy axis and can leave it only as complex-conjugate pairs~\cite{NockelStone1994,Lee1999}. Consistently, the phase of $t$ jumps by $\pi$ across each of the two zeros at $-V$, at $E\approx0.019$ and $E_0\approx0.084$~meV, where $|t|<10^{-10}$.

We track the zeros as a function of bias with a rational (AAA) approximation~\cite{Nakatsukasa2018} of $t(E)$, computed on the real axis for $-0.4\leq E\leq0.8$~meV. As the reverse bias is reduced, the two zeros inside the window approach each other, collide at $eV\approx-0.27$~meV, and leave the real axis as a complex-conjugate pair. At $V=0$ and at $+V$, the pair lies about $0.1$~meV off the real axis, where it remains for all forward biases we considered, up to about $0.5$~meV. Since the two zeros at $-V$ lie only about $0.07$~meV apart, $\mathcal T(E,-V)$ stays below $0.05$ for $0\leq E\leq e|V|/2$ and below $0.3$ for $-e|V|/2\leq E<0$, which gives $G(-V)\approx0.1G_0$. A field component $\Delta_y\sigma_y\tau_0$, which breaks $\Theta$, lifts the zeros, but $\Delta G/G_0\approx0.82$ ($0.69$) remains for $\Delta_y=0.01\Delta$ ($0.1\Delta$).

\subsection{Stability}
\label{sec:stability}

The maps of Fig.~3(d--f) vary two parameters at a time on an $81\times81$ grid of fractional deviations of up to $\pm10\%$, while all other parameters are held at Eq.~\eqref{eq:par-two-strong}. Along the axes of the maps, i.e., varying one parameter at a time, $\Delta G/G_0\geq0.8$ holds within about $\pm1\%$ of $\Delta$, $\theta$, and $U_0$, $\pm2\%$ of $d$, and $\pm4\%$ of $\mu$, and from $-10\%$ to $+5\%$ of $U_c$. The weaker condition $\Delta G/G_0\geq0.5$ holds over about $\pm3\%$ in $U_0$, $\pm6\%$ in $\Delta$ and $\theta$, and the full $\pm10\%$ range in $U_c$, $d$, and $\mu$.

\subsection{Resonant 1D device}

The same mechanism operates in the 1D model of Eq.~(6), where the upper spin-split band $E_+(k)$ plays the role of the closed channel. For $\alpha_R=0$ and $M=0$, the attractive scalar gates bind states of the upper band at $E\approx-0.02$ and $1.0$~meV, which the spin--orbit coupling and the magnetic scatterer couple to the open lower band. At the parameters of Eq.~\eqref{eq:par-one-strong}, we find $G(+V)\approx0.89G_0$ and $G(-V)\approx0.52G_0$, such that $\Delta G/G_0\approx0.37$ [see Fig.~\ref{fig:lambda}(b)]. The transmission zero at reverse bias lies at $E\approx-0.19$~meV, outside the bias window, such that the reverse current is suppressed less than in the quasi-1D device.

\section{Numerical parameters}
\label{sec:parameters}

We quote parameters to at most four significant figures, which reproduces all results to the precision given. These digits are needed only to reproduce the quoted numbers. The response itself persists under deviations of a few percent in each parameter (Sec.~\ref{sec:stability}). All calculations use $m^*=0.014m_e$, $\alpha_R=0.5\,\mathrm{eV\,\mathring A}=50\,\mathrm{meV\,nm}$, and $T=25\,\mathrm{mK}$. The quasi-1D calculations additionally use $W_y=200\,\mathrm{nm}$, for which Eq.~\eqref{eq:hardwall} gives
\begin{equation}
    \Delta_\perp=2.014\,\mathrm{meV},\qquad
    \lambda_0=0.6667\,\mathrm{meV}.
    \label{eq:common-two-band-values}
\end{equation}
The Zeeman energy $\Delta$ is the coefficient of $\hat{\bm n}\cdot\bm\sigma$, i.e., $\Delta=|g|\mu_BB/2$ for a field $B$ along $\hat{\bm n}$, with $\Delta_x=\Delta\sin\theta$ and $\Delta_z=\Delta\cos\theta$ in the quasi-1D model.

For the weak-scattering 1D device of Fig.~2(a--c), the operating point is
\begin{equation}
    \begin{gathered}
    \Delta_x=0.49\,\mathrm{meV},\quad \mu=-0.32\,\mathrm{meV},\quad U_0=2.85\,\mathrm{meV\,nm},\quad M=-4.76\,\mathrm{meV\,nm},\\
    d=120\,\mathrm{nm},\quad e|V|=0.08\,\mathrm{meV},\quad \Gamma eV/U_0=0.91,
    \end{gathered}
    \label{eq:par-one-weak}
\end{equation}
such that $\Gamma=32.4\,\mathrm{nm}$ and $k_Fd/\pi=0.54$, where $k_F$ is the value of $k_E$ at $E=0$. The exact values are $G(+V)\approx0.999G_0$, $G(-V)\approx0.955G_0$, and $\Delta G/G_0\approx0.045$, whereas the third-order Born result gives $\Delta G/G_0\approx0.046$.

For the weak-scattering quasi-1D device of Fig.~2(d), the operating point is
\begin{equation}
    \begin{gathered}
    \Delta=1.007\,\mathrm{meV},\quad \theta=52.75^\circ,\quad \mu=0.1749\,\mathrm{meV},\\
    U_0=14.59\,\mathrm{meV\,nm},\quad U_c=-29.18\,\mathrm{meV\,nm},\quad d=100\,\mathrm{nm},\quad e|V|=0.05\,\mathrm{meV},
    \end{gathered}
    \label{eq:par-two-weak}
\end{equation}
with $\Gamma eV/U_0=1$, i.e., $\Gamma=292\,\mathrm{nm}$, such that the right gate vanishes at $+V$, and with $U_c=-2U_0$ and $k_Fd/\pi=0.86$. The exact calculation gives $G(+V)\approx0.910G_0$, $G(-V)\approx0.908G_0$, and $\Delta G/G_0\approx1.6\times10^{-3}$. To make the interference mechanism visible, these two devices use much larger couplings than the resonant ones.

For the resonant quasi-1D device of Fig.~3, the operating point is
\begin{equation}
    \begin{gathered}
    \Delta=1.273\,\mathrm{meV},\quad \theta=77.98^\circ,\quad \mu=-1.032\,\mathrm{meV},\\
    U_0=-155.2\,\mathrm{meV\,nm},\quad U_c=-624.7\,\mathrm{meV\,nm},\quad d=100\,\mathrm{nm},\quad \Gamma=5\,\mathrm{nm},\quad e|V|=0.3016\,\mathrm{meV}.
    \end{gathered}
    \label{eq:par-two-strong}
\end{equation}
The bias changes either outer gate strength by only $|\Gamma eV|\approx1.5\,\mathrm{meV\,nm}$, or about $1\%$ of $|U_0|$. Nonetheless, the two conductances are $G(+V)\approx0.92G_0$ and $G(-V)\approx0.096G_0$, such that $\Delta G/G_0=0.82$, $\eta=0.81$, and $G(+V)/G(-V)\approx10$. These parameters are realistic for InSb nanowires, since $\Delta$ corresponds to a field of $0.88$~T for $|g|=50$, within the range measured in InSb nanowire quantum dots~\cite{Nilsson2009}. For a gate profile of width $20$~nm, the outer and central gate strengths correspond to potential wells about $8$ and $31$~meV deep, respectively. The coupling $\Gamma$ depends on the electrostatics of a given device and is the least constrained parameter. With all other parameters fixed and only the bias retuned, the largest $\Delta G/G_0$ is about $0.6$ for $\Gamma=2.5$~nm and $0.3$ for $\Gamma=1$~nm.

For the resonant 1D device of Fig.~\ref{fig:lambda}(b), the parameters are
\begin{equation}
    \begin{gathered}
    \Delta_x=1.390\,\mathrm{meV},\quad \mu=-1.189\,\mathrm{meV},\quad U_0=-154.4\,\mathrm{meV\,nm},\quad M=33.23\,\mathrm{meV\,nm},\\
    d=78.51\,\mathrm{nm},\quad \Gamma=5\,\mathrm{nm},\quad e|V|=0.3\,\mathrm{meV}.
    \end{gathered}
    \label{eq:par-one-strong}
\end{equation}

\section{Comparison with a single gate}
\label{sec:single}

\begin{figure}[t]
    \centering
    \includegraphics{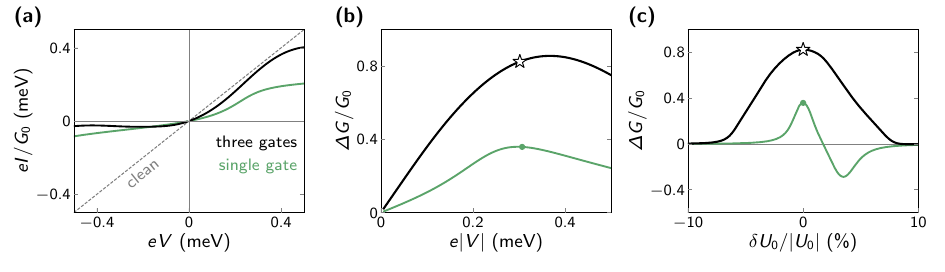}
    \caption{\textbf{Comparison with a single-gate rectifier.}
    (a) Current--voltage characteristics of the optimized single-gate device (green) and of the resonant three-gate device (black), with the clean wire shown dashed. (b) Diode effectiveness $\Delta G/G_0$ as a function of bias magnitude. (c) $\Delta G/G_0$ at the operating bias as a function of the fractional detuning of $U_0$, the strength of the single gate or of the two outer gates, with the central gate held fixed. The star and the dot mark the operating points of Eqs.~\eqref{eq:par-two-strong} and~\eqref{eq:single-contact-point}, respectively. The three-gate device produces both a substantially larger response and a much broader range of strong rectification.}
    \label{fig:device_comparison}
\end{figure}

For comparison, we optimized a single scalar point-like gate with $U(\pm V)=U_0\pm\Gamma eV$ in the same quasi-1D model, with the same $m^*$, $\alpha_R$, $W_y$, $T$, and coupling $\Gamma=5\,\mathrm{nm}$. The independently optimized point has
\begin{equation}
    \begin{gathered}
    \Delta=1.341\,\mathrm{meV},\quad \theta=80.23^\circ,\quad \mu=-1.240\,\mathrm{meV},\\
    U_0=-221.0\,\mathrm{meV\,nm},\quad e|V|=0.3066\,\mathrm{meV},
    \end{gathered}
    \label{eq:single-contact-point}
\end{equation}
where $|\Gamma eV|\approx1.5\,\mathrm{meV\,nm}$, about $0.7\%$ of $|U_0|$. Tuning to the single-gate resonance gives $\Delta G/G_0\approx0.36$. Even after independent optimization, this response is less than half of the three-gate value $\Delta G/G_0=0.82$ (see Fig.~\ref{fig:device_comparison}). It is also much more sensitive to $U_0$, since a $1\%$ detuning reduces the single-gate response to $\Delta G/G_0\simeq0.17$, whereas the three-gate device retains $\Delta G/G_0\simeq0.80$. At $2\%$ detuning, the former falls below $0.06$ and reverses sign on one side of the resonance, whereas the latter remains above $0.70$. The position of the gate drops out of the transmission, whereas its displacement from the device center produces the unequal capacitances encoded by $\Gamma$.

\bibliography{rectification_SM}